\documentclass[12pt, draftclsnofoot, onecolumn]{IEEEtran}

\ifCLASSINFOpdf
\else
\fi
\usepackage{url} 

\usepackage{dblfloatfix}    
\usepackage{amsfonts}
\usepackage{amsthm}
\usepackage{amssymb}
\usepackage{mathtools}
\usepackage{amsmath} 

\theoremstyle{definition}

\usepackage{mathtools, cuted} 
\usepackage{cuted}

\usepackage{url} 
\usepackage{blindtext, graphicx}

\usepackage{float} 
\usepackage{subcaption} 

\usepackage{multirow}
\usepackage{array}

\usepackage[numbers,sort&compress]{natbib}

\usepackage{booktabs}
\usepackage{tabularx}

\usepackage{array,multirow}

\usepackage{pbox}

\usepackage{supertabular}

\usepackage{dblfloatfix}
\usepackage{enumerate} 

\usepackage[linesnumbered,ruled,vlined]{algorithm2e}

\usepackage[table]{xcolor} 

\usepackage{amssymb} 
\usepackage{amsfonts}
\usepackage{bbm}

\usepackage{bm}

\DeclareMathAlphabet{\mathcal}{OMS}{cmsy}{m}{n}

\newcommand*\diff{\mathop{}\!\mathrm{d}}

\usepackage[normalem]{ulem} 

\begin{document}

\title{Analysis and Design of Distributed MIMO Networks with a Wireless Fronthaul}

\author{Hussein~A.~Ammar\IEEEauthorrefmark{1},~\IEEEmembership{Student Member,~IEEE}, Raviraj~Adve\IEEEauthorrefmark{1},~\IEEEmembership{Fellow,~IEEE},
	 Shahram~Shahbazpanahi\IEEEauthorrefmark{2}\IEEEauthorrefmark{1},~\IEEEmembership{Senior Member,~IEEE},
	and~Gary~Boudreau\IEEEauthorrefmark{3},~\IEEEmembership{Senior Member,~IEEE}
	\thanks{This work was supported in part by the Natural Sciences and Engineering Research Council (NSERC) of Canada and in part by Ericsson Canada.}
	\thanks{
		\IEEEauthorrefmark{1}H. A. Ammar and R. Adve are with the Edward S. Rogers Sr. Department of Electrical and Computer Engineering, University of Toronto, Toronto, ON M5S 3G4, Canada (e-mail: ammarhus@ece.utoronto.ca; rsadve@comm.utoronto.ca).
	}
	\thanks{
		\IEEEauthorrefmark{2}S. Shahbazpanahi is with the Department of Electrical, Computer, and Software Engineering, University of Ontario Institute of Technology, Oshawa, ON L1H 7K4, Canada. He also holds a Status-Only position with the Edward S. Rogers Sr. Department of Electrical and Computer Engineering, University of Toronto.
	}
	\thanks{
		\IEEEauthorrefmark{3}G. Boudreau is with Ericsson Canada, Ottawa, ON K2K 2V6, Canada.
	}
}


\maketitle 

\makeatletter
\def\tagform@#1{\maketag@@@{\normalsize(#1)\@@italiccorr}}
\makeatother

\vspace{-3em}
\begin{abstract}
	We consider the analysis and design of distributed wireless networks wherein remote radio heads (RRHs) coordinate transmissions to serve multiple users on the same resource block (RB). Specifically, we analyze two possible multiple-input multiple-output wireless fronthaul solutions: multicast and zero forcing (ZF) beamforming. We develop a statistical model for the fronthaul rate and, coupled with an analysis of the user access rate, we optimize the placement of the RRHs. This model allows us to formulate the location optimization problem with a statistical constraint on fronthaul outage. Our results are cautionary, showing that the fronthaul requires considerable bandwidth to enable joint service to users. This requirement can be relaxed by serving a low number of users on the same RB. Additionally, we show that, with a fixed number of antennas, for the multicast fronthaul, it is prudent to concentrate these antennas on a few RRHs. However, for the ZF beamforming fronthaul, it is better to distribute the antennas on more RRHs. For the parameters chosen, using a ZF beamforming fronthaul improves the typical access rate by approximately $8\%$ compared to multicast. Crucially, our work quantifies the effect of these fronthaul solutions and provides an effective tool for the design of distributed networks.
\end{abstract} 

\IEEEpeerreviewmaketitle
\section{Introduction}
Distributed multiple-input multiple-output (MIMO) systems provide a paradigm shift for wireless networks through enhanced connectivity, reliability guarantees, interference management, and macro-diversity. This shift can be achieved through the distributed architecture of the next-generation NodeB (gNB) which comprises a central unit (gNB-CU) connected to distributed units (gNB-DUs), through the F1 interface~\cite{3GPP:TR21.915}. The implementation  proposed for the distributed architecture supports different upper- and lower-layer functional splits. Of relevance to this study is allowing the network transmitters to jointly serve the users through a joint construction of beamformers. Thus, by assuming a lower-layer functional split of the protocol layers (performed at the F2 interface), we use the term CU to represent the node that hosts both the functionalities of the gNB-CU and the construction of the precoding matrices to be used by the network transmitters, and we use the term remote radio heads (RRHs) to refer to the units that are distributed throughout the network and host simple physical layer functionalities needed to serve the users. The interface between the CU and the RRH is referred to as the fronthaul. Hence, in a nutshell, the CU controls the transmissions to and from the RRHs, while the RRHs coordinate transmissions, e.g., through joint transmission, and enable multiuser communication on the same time-frequency resource block (RB) through a distributed antenna~system.

The fronthaul has a huge impact on the deployed locations of the RRHs. In many applications of interest, providing a wired connection between the CU and the RRHs is impractical - this may be due to the location of RRHs or to retain the ability to quickly build up/tear down a network. This highlights the importance of \emph{wireless fronthaul}. However, using wireless fronthaul imposes a coverage constraint that restricts the locations of the RRHs. It also establishes difficulties in designing the network because the communication on the fronthaul (CU-to-RRH) and on the access (RRH-to-user) channels are \emph{dependent}. In turn, the fronthaul scheme and the deployed locations of the RRHs affect network performance, thus both must be chosen carefully~\cite{EhsanBackhaul}.  

In this paper, we provide both analytical tools and design techniques to optimize the locations of the RRHs in distributed networks with fronthaul constraints and with a non-homogeneous spatial user density. To the best of our knowledge, such analysis has not been attempted, with previous works ignoring the fronthaul~\cite{8529184} or using a simple fronthaul scheme~\cite{ammarWirelessbackahulConf}. We study two different wireless fronthaul solutions exploiting MIMO communication techniques. The first uses multicast beamforming (BF), where each CU transmits a single message to all the RRHs it controls. In this case, we extend the (asymptotically optimal) closed-form beamformer expressions for a single receive antenna~\cite{6824794} to the case of multiple receive antennas per RRH. The second scheme employs zero forcing (ZF) beamforming in the fronthaul.

The random fading in the fronthaul channels leads to a randomness in fronthaul capacity. Hence, the analysis of these links requires a statistical characterization, which is, unfortunately, mathematically intractable. We therefore model the signal and interference powers on these links as random~\cite{8449213}. In~\cite{ZFBFLetter9141340}, we use the Kolmogorov–Smirnov (KS) test to show that a gamma distribution best matches the distribution of the signal and interference powers while also providing closed-form expressions that can be optimized. One contribution of this paper is to derive closed-form expressions for the mean and variance of these power levels, under both fronthaul schemes considered, required to fit the gamma random variable (RV). In turn, we obtain closed-form expressions for the fronthaul outage. Crucially, this allows us to formulate an optimization problem for the locations of RRHs while accounting for the randomness in the fronthaul.

The deployed locations for the RRHs play a crucial role in maximizing the access rate. One of the earliest studies on RRH location optimization is~\cite{4939337}, which considers a network with a single cell, noise-limited environment, and no restrictions on the fronthaul capacity. The study in~\cite{7996634} assumes that there are two types of RRHs; a primitive one and an RRH with layer 1 functions, and then formulates an objective function to select the sites (from available ones) and the number of nodes to deploy for each type. The authors assume a fixed limited-capacity fronthaul. The optimal placement of relays in a non-cooperative cellular network is studied in~\cite{6702841}. The authors model the user traffic using queueing theory, and they consider time division between the fronthaul and access channels of relaying nodes for in-band transmissions, and frequency division for out-of-band transmissions. The main results show that in-band relaying provides lower performance compared to the latter. The investigation in~\cite{8529184} considers the placement of the RRHs in the presence of dedicated fronthaul links with high capacity and low latency. Furthermore, the study in~\cite{tonini2019cost} places the RRHs in selected locations by formulating an integer linear program that minimizes the deployment cost while guaranteeing a capacity requirement.

The cloud radio access network (C-RAN) literature has many suggested transport standards for the fronthaul. The most famous one is the common public radio interface (CPRI) specification which is a serial fronthaul interface~\cite{7402275}. This specification defines a functional split so that almost no digital processing functions are required at the RRHs, making them small and cheap. CPRI also separates the traffic into three logical connections: data plane, control and management plane, and synchronization and timing. A wireless fronthaul has been analyzed in~\cite{8283646, 8830409, 7925528, 7358079} for C-RAN. Specifically, \cite{8283646} optimizes the beamformers on the access and fronthaul links using a weighted sum rate problem that is approximated by a sequence of convex problems. While~\cite{8830409} considers the joint optimization of the quantization noise covariance and the time allocation for the access and fronthaul links so that the rate is maximized. Furthermore,~\cite{7925528} minimizes the transmit power under an SINR constraint while using a hybrid wireless and optical fronthaul network, and \cite{7358079} minimizes the transmit power while guaranteeing a predefined system block error rate. The architecture of these works is based on a single centralized baseband unit (BBU) pool, which is different from our network scheme that uses multiple CUs which is consistent with the distributed architecture of gNBs~\cite{3GPP:TR21.915}. In this regard, using a wireless fronthaul with a centralized BBU pool is not appropriate due to the possible large distances between the BBU and the RRHs. Furthermore, these studies optimize the beamformers, but do not optimize the locations of the RRHs when constrained by a limited wireless fronthaul. Thus, they are not meant to be used to design distributed MIMO networks before the deployment of the RRHs.

Fronthaul links with a preset fixed capacity have been analyzed in the literature. For example, the authors of~\cite{7952837} analyze joint beamforming and cooperative cluster selection to maximize the sum rate. Similarly, the authors of~\cite{6571307} investigate soft and hard switching between coordinated multipoint (CoMP) joint transmission (JT) and coordinated beamforming (CB), where soft switching corresponds to using CoMP-CB for private data and CoMP-JT for common data. Additionally, the study in~\cite{7581201} considers the joint optimization of resource allocation, while that in~\cite{6920005} considers user scheduling and base stations clustering with fronthaul resource constraints.

In the literature, no studies have been conducted to optimize the location of the RRHs under a wireless fronthaul solution. Moreover, most works targeting a limited fronthaul capacity are based on using a fixed capacity fronthaul. This means that the fronthaul capacity does not change with the problem variables such as the location of the RRHs. We note that this problem, under the scenarios considered in this paper, is still not studied even for other distributed MIMO architectures~\cite{9113273}, such as the user-centric cell-free [massive] MIMO scheme~\cite{ammar2021user}.

Our study fills a gap in the literature by studying the effect of a wireless fronthaul on the deployed locations of the RRHs. The contributions of our work can be summarized as follows:
\begin{itemize}
	\item We provide a statistical analysis for two MIMO wireless fronthaul solutions, multicasting and ZF, in a multi-cell network, in turn providing a statistical characterization of fronthaul outage. During this process, we derive accurate closed-form approximations for the mean and variance of the desired signal and interference power levels. The statistical analysis is crucial because the placement of the RRHs is a long-term problem, i.e., it should not be dependent on instantaneous fluctuations of the channels.
	\item Equipped with the closed-form expressions, we develop an optimization problem that maximizes the access capacity by optimizing the locations of the RRHs given a statistical fronthaul outage constraint and environments with non-homogeneous spatial density for the users. We then propose an effective algorithm to solve this optimization problem. Compared to the literature our study provides a contribution by targeting both RRH placement and wireless fronthaul limited capacity solutions in the same work, which has not been done before. Note that RRH placement under wireless fronthaul is a complicated task because any change in the locations of the RRHs impacts the fronthaul capacity.
	\item We present results to show the effect of the two fronthaul schemes on the RRH deployment decisions, and we investigate the trade-offs between important network parameters, such as the bandwidth division between the access channel and fronthaul, the number of antennas on the RRHs, the number of RRHs, and the number of users to be served on the same RB.
	\item We show that for a ZF fronthaul, it is better to distribute the antennas on as many RRHs as possible. In contrast, for a multicast fronthaul scheme it is better to limit the number of RRHs per cell. For the parameters chosen, compared to using multicast, using ZF in the fronthaul can improve typical access rates by 8\%.
\end{itemize}

\noindent \textit{Notations Used}: A bold lower case letter ${\bf a}$ represents a vector, while a bold upper case letter ${\bf A}$ represents a matrix. The term $[{\bf a}]_i$ is the $i$th entry of a vector ${\bf a}$, $[{\bf A}]_{ij}$ is the $(i,j)$th entry of a matrix ${\bf A}$, while $[{\bf A}]_{.j}$ is the $j$th column of a matrix ${\bf A}$. The operators $(\cdot)^{-1}$, $(\cdot)^T$, $(\cdot)^*$, and $(\cdot)^H$ denote the matrix inverse, transpose, conjugate, and conjugate transpose, respectively. Additionally, $\mathbb{E}\{\cdot\}$, $\mathtt{Var}\{\cdot\}$, and $\mathbb{P}\{\cdot\}$ represent the expectation, variance, and probability operators, respectively. Unless specified, expectation is over the small-scale fading. $\text{tr}\{\cdot\}$ is the trace of a matrix, $\|\cdot\|$ and $|\cdot|$ are the vector and scalar norms, $\lceil \cdot \rceil$ is the ceiling operator, and ${\bf I}_m$ is the $m$-dimensional identity matrix. The terms $\mathbb{R}^m$, $\mathbb{C}^m$, and $\mathbb{C}^{m\times n}$ are the set of real and complex $m\times 1$ column vectors, and complex $m\times n$ matrix, respectively. $\mathbf{x} \sim \mathcal{CN}(\mathbf{m},\mathbf{R})$ indicates that $\mathbf{x}$ is a complex Gaussian random vector with mean $\mathbf{m}$ and covariance $\mathbf{R}$.

\noindent \textit{Outline of Paper}: The rest of the paper is organized as follows. Section~\ref{sec:system_model} defines the system model of the paper and presents important preliminaries needed to formulate our RRHs deployment problem. Section~\ref{sec:multicast_singleRXAnt} presents the multicast schemes for the fronthaul and formulates our optimization problem. Section~\ref{sec:fronthaul_ZFBF} presents a similar analysis, but for the ZF beamforming fronthaul scheme. Section~\ref{sec:results} presents numerical results illustrating the accuracy of our analysis and results of the optimization. Finally, Section~\ref{sec:conclusion} concludes this study.
\begin{figure}[t]
	\centering
	\includegraphics[width=0.6\textwidth]{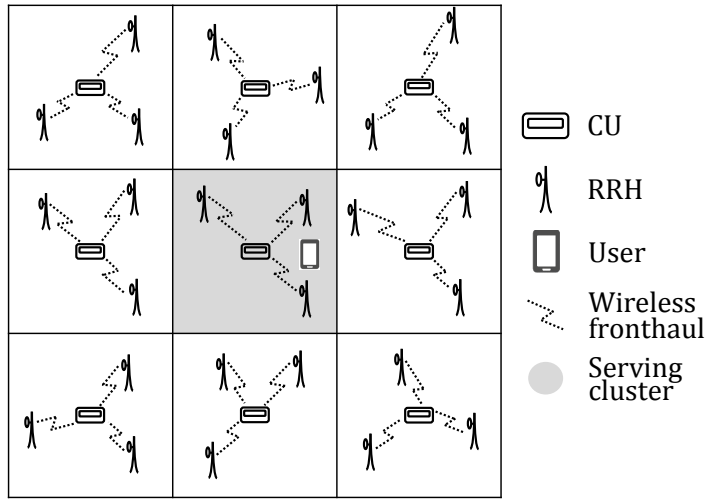}
	\caption{A typical network scheme.}
	\vspace{-1.2em}
	\label{fig:NetDisjoint}
\end{figure}

\section{System Model}\label{sec:system_model}
We consider a wireless network comprising $Q$ cells, each containing one CU assumed located at the cell center (though our analysis is valid for any CU location inside the cell) and connected to $N$ RRHs distributed throughout the cell. The CUs are the endpoints of the core network and are connected to the RRHs through wireless channels, hence providing a fronthaul capacity that depends on the fronthaul scheme, channel realizations, and the locations of the RRHs. Each CU, RRH, and user is equipped with $M_{\rm c}$, $M$ and $1$ antenna, respectively. Data transmissions on the fronthaul and access channels are separated through frequency division. We assume ideal sub-division of bandwidth between the channels, but a frequency overhead can be easily included. \label{page:perfectCSI}To focus on network design, we assume that perfect channel state information (CSI) is available as needed, and we analyze the data plane, hence, under a fixed JT service, our analysis is not affected by any differences in the functional split between the CU and the RRH. Furthermore, each CU has access to the data of all users within its cell.

We use a cell-centric (or disjoint clustering) network model~\cite{PDPUsercentricVsDisjoint8969384}: the $N$ RRHs in cell $q$ form a cluster $\mathcal{D}_q$ to serve $K$ users within the same cell on the same RB. We denote the set of users served across the network as 
${\mathcal{U}}=\{\mathcal{U}_1,\dots,\mathcal{U}_Q\}$, where ${\mathcal{U}}_q\subset {\mathcal{U}}$ corresponds to $K$ users served in cell $q$. These users can be non-uniformly distributed due to the existence of hotspots, i.e., dense areas. We aim to optimize the locations of the RRHs to maximize the access channel rates. We consider square cells to use the Cartesian coordinate system in our problem, however, other shapes like circular cells can be used by implicit conversion to polar coordinates when needed. In Fig.~\ref{fig:NetDisjoint}, we depict a typical network. With a non-uniform spatial distribution for users, the RRHs need to be also deployed in a non-uniform pattern to obtain good performance; importantly, the wireless fronthaul will restrict the locations of the RRHs due to possible fronthaul outage. We focus on non line-of-sight (NLoS) fronthaul because maintaining LoS communication cannot be guaranteed when optimizing the locations of the RRHs, especially when $N$ is large. This is because, unlike traditional base stations, RRHs may not be placed at a height. We note that, with considerably increased complexity, our analysis can be extended to Rician fading~\cite{ammarWirelessbackahulConf}.

\subsection{Signal Model on the Access Channel}
A user $k$ located in cell $q$, i.e., $k\in {\mathcal{U}}_q$ and served by the RRH cluster $\mathcal{D}_{q}$, receives the signal
\begin{align}\label{eq:rx_access}
\resizebox{0.9\textwidth}{!}
{$\displaystyle
	r_{qk} = \displaystyle \underbrace{\sum_{n\in \mathcal{D}_{q}}{\bf h}_{qnk}^H {\bf w}_{qnk} s_{qk}}_{\text{desired signal}}
	+\underbrace{\sum_{n\in\mathcal{D}_{q},k'\in {\mathcal{U}}_q,k'\ne k}{\bf h}_{qnk}^H {\bf w}_{qnk'} s_{qk'}}_{\substack{\text{intra-cluster}\text{ interference}}}
	+\underbrace{\sum_{n'\in \mathcal{D}_{q'} ,q' \ne q, k' \in {\mathcal{U}}_{q'}} {\bf h}_{q'n'k}^H {\bf w}_{q'n'k'} s_{q'k'}}_{\substack{\text{inter-cluster}\text{ interference}}}
	+ z_{qk}
	$}, 
\end{align}
where $s_{qk}\in\mathbb{C}$ is the unit-energy symbol sent to user $k$ by the serving RRHs. The vector ${\bf h}_{qnk}^*\in\mathbb{C}^{M}$ denotes the channel between RRH $n$ in cell $q$ and user $k$. We model ${\bf h}_{qnk}$~as
\begin{align}\label{eq:accessChannel}
{\bf h}_{qnk}=\sqrt{\beta_{qnk}\ell_k(x_{qn},y_{qn})}{\bf g}_{qnk},
\end{align}
where ${\bf g}_{qnk}\sim\mathcal{CN}({\bf 0},{\bf I}_M)$ represents the small-scale fading and \sloppy{$\ell_k(x_{qn},y_{qn}) = \left(1+d_{qnk}/d_0\right)^{-\alpha}$} represents the path loss, with $d_{qnk}=\sqrt{(x_{qn} - \widetilde{x}_k)^2 + (y_{qn} - \widetilde{y}_k)^2}$ the distance between RRH $n \in \mathcal{D}_q$ and user $k$, $d_0$ the reference distance, $\alpha$ the path loss exponent, and $\beta_{qnk}$ denotes the shadowing. The terms $(x_{qn},y_{qn})$ denote the x and y coordinates of RRH $n$, while the vector ${\bf w}_{qnk}\in\mathbb{C}^M$ denotes the precoding vector used by RRH $n$ for user $k$. Finally, $z_{qk}\sim \mathcal{CN}(0,\sigma_z^2)$ denotes the additive white Gaussian noise (AWGN) with average power~$\sigma^2_z$.

We define the concatenation of the channel matrix for all the users in cell $q$ as ${\bf H}_{q}^* \triangleq [{\bf h}_{q1}^*\dots{\bf h}_{qK}^*]\in\mathbb{C}^{NM\times K}$, with ${\bf h}_{qk}^*=[{\bf h}_{q1k}^H\dots{\bf h}_{qNk}^H]^T$ being the length-$NM$ concatenation of the channels between the $N$ RRHs in cell $q$ and user $k$. We can write ${\bf h}_{qk} \triangleq {\bf D}_{qk}^{1/2}{\bf g}_{qk}$, where ${\bf D}_{qk}\in \mathbb{C}^{NM\times NM}$ is a diagonal matrix representing the large-scale fading between all the RRHs in cell $q$ and user $k$. Hence,  
$[{\bf D}_{qk}]_{mm}=\beta_{qnk}\ell_k(x_{qn},y_{qn})$ where $n=\lceil{m/M}\rceil$, for $m=\{1,\dots,NM\}$. Correspondingly, ${\bf g}_{qk}\in \mathbb{C}^{NM}$ represents the vector of the small-scale fading coefficients between the antennas of the RRHs in cell $q$ and user $k$.

The precoding matrix designed for cell $q$ is defined as ${\bf W}_{q}=[{\bf v}_{q1} \dots {\bf v}_{qK}]\in\mathbb{C}^{NM\times K}$ with ${\bf v}_{qk}=[{\bf w}_{q1k}^T \dots {\bf w}_{qNk}^T]^T\in\mathbb{C}^{NM}$, where ${\bf w}_{qnk} \in \mathbb{C}^M$ is defined above. We use ZF linear precoding on the access channel. With the CSI known, the precoding matrix is formed as
\begin{align}\label{eq:beamformerAccess}
{\bf W}_{q}={\bf \widetilde{W}}_{q}\bm{\mu}_q 
= {\bf H}_{q}\left({\bf H}_{q}^H{\bf H}_{q}\right)^{-1} \bm{\mu}_q,
\end{align}
where $\bm{\mu}_q \in \mathbb{C}^{K \times K}$ allows us to satisfy the power constraint. As done in~\cite{6823643, vectorNormalization6477575}, we use a \emph{vector normalization}, allowing for a fair comparison across network configurations. Hence, $\bm{\mu}_q$ is a diagonal matrix such that $\mathbb{E}\left\{\text{tr}\left\{ {\bf W}_{q} {\bf W}_{q}^H\right\}\right\} = p$; its $k^{th}$ diagonal entry is defined as
\begin{align}
\mu_{qk} = \left[\bm{\mu}_q\right]_{kk} = \sqrt{p K^{-1} \mathbb{E}\{\|{\bf \widetilde{v}}_{qk}\|^2\}^{-1}} \label{eq:mu_qk}
\end{align}
with ${\bf \widetilde{v}}_{qk}=\left[{\bf \widetilde{W}}_{q}\right]_{.k} = \left[{\bf H}_{q}\left({\bf H}_{q}^H{\bf H}_{q}\right)^{-1}\right]_{.k}$, i.e., the $k^{th}$ column of ${\bf \widetilde{W}}_{q}$. Note that we assume the matrix ${\bf H}_{q}$ is full-row rank, an assumption that is valid with probability one~\cite{6823643}.

\subsection{Spectral Efficiency of Access Channel}
The ZF beamformer used for the access channel is constructed for each cell $q$, thereby canceling intra-cluster interference. Based on the normalization in~\eqref{eq:mu_qk}, the average signal power received at user $k \in \mathcal{U}_q$ in cell $q$ is $\mu_{qk}^2$. Under our assumption of perfect CSI, one lower bound on the achievable spectral efficiency, measured in nats/s/Hz, is~\cite{1193803}
\begin{align}
R_{qk}^{\rm{(a)}}
=
\log
\left(
\rule{0cm}{0.8cm}\right.
1+\frac{ \mu_{qk}^2} { \sum_{n'\in \mathcal{D}_{q'}, q' \ne q,k'\in {\mathcal{U}}_{q'}} \mathbb{E}\left\{ \mid {\bf h}_{q'n'k}^H {\bf w}_{q'n'k'}\mid^2 \right\} + \sigma_z^2} 
\left.\rule{0cm}{0.8cm}
\right)
.
\label{eq:AccessRate_1}
\end{align}
We note that, through the channel powers, this spectral efficiency is a function of the distances between the RRHs found in the network and user $k$, (later on the RRH locations which will be our optimization variables). The expectation in~\eqref{eq:AccessRate_1} is over the small-scale fading realizations \textit{and} user locations.

Since user locations are unknown at the system design stage, we treat user locations as random, and hence we assume that their statistics are represented through a spatial traffic distribution (assumed known). This distribution inside a specific network area can be constructed using passive monitoring of mobile users in the network regions, which can be located through location areas, sectors, cells, etc. Then the hotspots in the network can be identified from the collected data. This distribution can also be built from very extensive and complex surveys performed for the network, such as identifying the locations of the crowded areas (e.g., stadiums, campuses, etc), and analyzing spatio-temporal activity of the users. All in all, our study is only concerned with the step of representing the spatial traffic distribution using a series of convex combinations, and hence we do not elaborate in detail on the methods of collecting the spatial statistics from an actual network; readers can refer to papers that specifically target this topic such as the survey in~\cite{survey7299258}.

Let the concatenated form of the RRH locations in the network be written as $\left[{\bf x},{\bf y}\right]$, where the x-coordinates are ${\bf x}=[{\bf x}_1^T,\dots,{\bf x}_q^T,\dots,\dots,{\bf x}_Q^T]^T$ (and similarly for ${\bf y}$). We use $[{\bf x}_q, {\bf y}_q]$ to refer to the locations of the RRHs in cell $q$, where ${\bf x}_q = [x_{q1}\ \dots\ x_{qN}]^T$. We also use $[{\bf x}_{-q},{\bf y}_{-q}]$ to refer to the locations of all the RRHs in the network \emph{other than} in cell $q$. Using this notation, this lower-bound of the achieved spectral efficiency on the access channel for a typical user $k$ served by the disjoint cluster $\mathcal{D}_q$ is~\cite{8529184}
\begin{align}\label{eq:spectralEfficiency_access}
R_{qk}^{\rm{(a)}}\left({\bf x}, {\bf y}\right)=
\log\Big(1+ \gamma_k\left({\bf x}_{-q},{\bf y}_{-q}\right)^{-1}\sum_{n\in \mathcal{D}_{q}}\beta_{qnk}\ell_k(x_{qn},y_{qn}) \Big)
\end{align}
with
\begin{align}
\gamma_k\left({\bf x}_{-q},{\bf y}_{-q}\right)=
\frac{NK}{(NM-K)\rho}\bigg(\displaystyle \frac{M \rho}{K} \sum_{q' \ne q}{I}^{(a)}_{q'k}({\bf x}_{q'},{\bf y}_{q'}) + 1\bigg)
\end{align}
Here, ${I}^{(a)}_{q'k}({\bf x}_{q'},{\bf y}_{q'})$ denotes an average inter-cluster interference (ICI) term. We begin with
\begin{align}
\check{I}^{(a)}_{q'k}({\bf x}_{q'},{\bf y}_{q'})
=
\sum_{n'\in\mathcal{D}_{q'}}\beta_{q'n'k}\ell_k(x_{q'n'},y_{q'n'})\sum_{j\in {\mathcal{U}}_{q'}}\frac{\beta_{q'n'j}\ell_j(x_{q'n'},y_{q'n'})}{\xi(q',j)},
\end{align}
where $\rho=\frac{p}{\sigma_z^2}$ and $\xi(q',j)=\text{tr}\{{\bf D}_{q'j}\}=M\sum_{m \in \mathcal{D}_{q'}}\beta_{q'mj}\ell_j(x_{q'm},y_{q'm})$. The RRHs in an interfering cell $q'$ contribute $\check{I}^{(a)}_{q'k}({\bf x}_{q'},{\bf y}_{q'})$ to the total inter-cell interference. This interference term depends on the random locations of the users in the interfering cells because the beamformers in each cell is designed based on the channels of these users. Let $(\widetilde{x}_k,\widetilde{y}_k)$ denote the location of typical user $k$. To obtain the average ICI on the access channel, we use a spatial traffic probability density function (PDF) $f_q(\widetilde{x}_k,\widetilde{y}_k)$ to model the user locations, resulting in
\begin{align}
& {I}^{(a)}_{q'k}({\bf x}_{q'},{\bf y}_{q'})  =
\mathbb{E}_{\widetilde{x}_j,\widetilde{y}_j}\{\check{I}^{(a)}_{q'k}({\bf x}_{q'},{\bf y}_{q'})\} =
\sum_{n'\in\mathcal{D}_{q'}}\beta_{q'n'k}\ell_k(x_{q'n'},y_{q'n'})K\mathbb{E}_{\widetilde{x}_j,\widetilde{y}_j}\left\{\frac{\beta_{q'n'j}\ell_j(x_{q'n'},y_{q'n'})}{\xi(q',j)}\right\}
\nonumber\\
&
\hspace*{0.15in}=
\sum_{n'\in\mathcal{D}_{q'}}\beta_{q'n'k}\ell_k(x_{q'n'},y_{q'n'}) K
\int \!\!  \int_{\widetilde{x}_j,\widetilde{y}_j\in \mathcal{B}_{q'}} \!\!
\frac{\beta_{q'n'j}\ell_j(x_{q'n'},y_{q'n'})}{\xi(q',j)} f_{q'}(\widetilde{x}_j,\widetilde{y}_j) \diff \widetilde{x}_j \diff \widetilde{y}_j,
\label{eq:AverageICI}
\end{align}
where $\mathcal{B}_{q'}$ denotes the boundary of cell $q'$. The PDF $f_q(\widetilde{x}_k,\widetilde{y}_k)$ can be constructed from a traffic survey, and it allows us to represent the density of users at any coordinate $(\widetilde{x}_k,\widetilde{y}_k)$. 
Later in the paper, we will elaborate on how to define $f_q(\widetilde{x}_k,\widetilde{y}_k)$. The derivation of~\eqref{eq:spectralEfficiency_access} from~\eqref{eq:AccessRate_1} can be found in~\cite[Appendix~A]{8529184}.

In summary,~\eqref{eq:spectralEfficiency_access}-\eqref{eq:AverageICI} provide a lower bound on the access rate to a typical user as a function of the locations of the RRHs. These rates are possible only if the fronthaul can sustain them. In the next two sections, we use our network model to analyze two possible fronthaul schemes: multicast and ZF beamforming. The choice of ZF and multicast beamforming is based on their tractability, in addition to their low computational complexity and low coordination requirements between the CUs compared to counterpart numerical beamforming techniques. Wireless multicasting is appropriate since all RRHs in a cell require the data of all $K$ users they serve, i.e., all $N$ RRHs are to receive the same data. However, as we will see, multicasting is effective only for a small number of $N$. For large $N$, we must use ZF beamforming in the fronthaul. 

\section{Multicast Fronthaul}\label{sec:multicast_singleRXAnt}
Wireless multicasting has been included in the LTE standards, known as evolved Multimedia Broadcast Multicast Service (eMBMS)~\cite{3GPP:TR36.743}, to provide an efficient way of delivering data that is common to the receivers. The main advantage of multicast arises from the fact that only one CU transmission is needed using one multicast beamformer. However, with multicast, the data rate is limited by the worst rate amongst all receivers.

The multicast signal received on the fronthaul at RRH $n$ from its serving CU $q$ is given by
\begin{align}
\bar{r}_{qn} =& \displaystyle \underbrace{{\bf \bar{h}}_{q,qn}^H {\bf \bar{w}}_{q} \bar{s}_{q}}_{\text{desired signal}}
+\underbrace{\sum_{q'\ne q}^{Q}{\bf \bar{h}}_{q',qn}^H {\bf \bar{w}}_{q'} \bar{s}_{q'}}_{\substack{\text{inter-cell}\text{ interference}}}
+ z_{qn},
\end{align}
where $\bar{s}_{q}$ is the (multicasted) data symbol with average power $p_{\rm c}$, and is common to all the RRHs in cell $q$. Here, $p_{\rm c}$ being the power budget of the CU; ${\bf \bar{w}}_{q} \in \mathbb{C}^{M_{\rm c}}$ is the multicast beamforming vector with $\mathbb{E}\{\|{\bf \bar{w}}_{q}\|^2\}=1$; ${\bf \bar{h}}_{q',qn}^* \in \mathbb{C}^{M_{\rm c}}$ denotes the fronthaul channel between the CU in cell $q'$ and RRH $n$ in cell $q$; finally $z_{qn}$ denotes the AWGN. Since all RRHs receive the same data using the common beamformer $\mathbf{\bar{w}}_q$, there is no intra-cell interference.


\subsection{Multicast Beamformer Design}
\subsubsection{Multicast for RRHs with single-receive antennas}\label{sec:multicast_singleAnt}
Since, in multicasting, a common message is transmitted to the RRHs in each cell, system performance is characterized by the performance of the RRH with the minimum SINR~\cite{6824794}. This implies that optimal beamformer design requires maximizing the minimum achieved SINR among the receivers in the same multicast group. This max-min problem has been studied in the literature and proven to be NP-hard~\cite{1634819}. This problem is defined as follows
	\begin{align}\label{eq:multicastBeamformer}
	&\underset{\mathcal{W}}{\max}\ \underset{n \in \mathcal{D}_q, \forall q}{\min}\quad 
	\frac{
		p_{\rm c}|{\bf \bar{h}}_{q,qn}^H {\bf \bar{w}}_{q}|^2
	}
	{
		p_{\rm c}\sum_{q'\ne q}^{Q} |{\bf \bar{h}}_{q',qn}^H {\bf \bar{w}}_{q'}|^2
		+ \sigma_z^2
	} 
	&\text{s.t.}\quad
	\|{\bf \bar{w}}_q\|^2 \le 1, \text{ for } q = 1, \dots, Q
	\end{align}
where $\mathcal{W} = [\mathbf{\bar{w}}_1^T, \mathbf{\bar{w}}_2^T, \dots, \mathbf{\bar{w}}_Q^T]^T$ is the concatenation of the multicast beamformers used by the CUs. Semidefinite Programming (SDP) has been found to be useful to solve such problems, where the beamformer is written as a weighted linear combination of the receivers' channels and the problem reduces to finding these sub-optimal weights. One proposed approach is to formulate this as a feasibility problem~\cite{8462123} solved by bisection followed by Gaussian randomization~\cite{6516905}.

Unfortunately, these proposed approaches suffer from three drawbacks: one, Gaussian randomization does not yield the optimal weights, i.e., we suffer a performance loss; second, solving the optimization problem is computationally complex. Third, and crucially for our purposes, the proposed approaches do not yield a closed-form solution for the beamformer, in turn, precluding a statistical analysis. As a viable alternative, we employ the asymptotically (in $M_{\rm c}$, the number of antennas at each CU) optimal beamformer proposed in~\cite{6824794} which has a closed-form expression. This beamformer relies on the fact that, asymptotically, the relevant channels become orthogonal. For a single receive antenna, this beamformer is given by~\cite{6824794}
\begin{align}\label{eq:asymptoticBeamformer_singleRxAnt}
{\bf \bar{w}}_{q} = \bar{\mu}_{q} \sum_{n=1}^{N} \frac{{\bf \bar{h}}_{q,qn}}{\bar{\ell}_q(x_{qn},y_{qn})},
\end{align}
where $\bar{\ell}_q(x_{qn},y_{qn})$ denotes path loss from CU $q$ to RRH $n$ in cell $q$ and $\bar{\mu}_{q}$ is a common factor that ensures $\mathbb{E}\{\|{\bf \bar{w}}_{q}\|^2\}=1$ and is given by
\[
\bar{\mu}_{q} = \frac{1}{\sqrt{ M_{\rm c} \sum_{n=1}^{N} \frac{1}{\bar{\ell}_q(x_{qn},y_{qn})}}}.
\]

\begin{figure}
	\centering
	\includegraphics[width=0.55\textwidth]{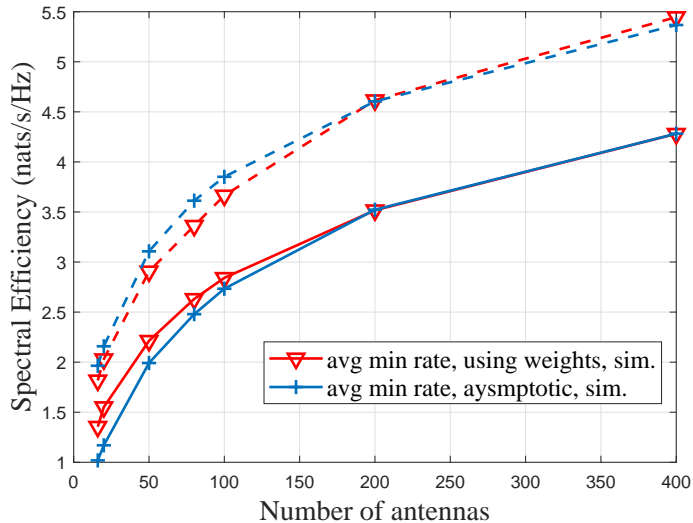}
	\caption{Multicast spectral efficiency versus number of CU antennas, $M_{\rm c}$, using SDP and asymptotic beamformer. $N=10$ (solid lines) and $N=5$ (dashed lines).}
	\vspace{-1.7em}
	\label{fig:SolidN10_dashedN5}
\end{figure}

To illustrate the validity of our approach, in Fig.~\ref{fig:SolidN10_dashedN5}, we compare the spectral efficiencies {at the fronthaul} obtained from the numerical SDP approach after randomization (the ``using weights'' legend) and from using the asymptotic multicast beamformer in~\eqref{eq:asymptoticBeamformer_singleRxAnt}. {We plot the results as a function of $M_{\rm c}$, the number of CU antennas, where the} parameters used are given in Table~\ref{table:sim_parameters} and described in Section~\ref{sec:results} later. For this plot, the RRHs are distributed randomly outside an exclusion zone of $100\ \rm{m}$ from the CU, and the rates are averaged over $1000$ random location realizations. As can be seen in this figure, the performance of the SDP approach after Gaussian randomization is not very different from that of the asymptotic beamformer. 

Given the agreement between the two approaches in Fig.~\ref{fig:SolidN10_dashedN5}, and the fact that the asymptotic beamformer provides a closed-form expression, we use~\eqref{eq:asymptoticBeamformer_singleRxAnt} for our further analysis. Importantly, all the work in the literature is for a single receive antenna, while, in our case, each receiver (the RRH) has $M$ antennas. We begin with analyzing the case of a single receive antenna and then extend our approach to the case of $M$ antennas.

It is worth noting that the beamformer in~\eqref{eq:asymptoticBeamformer_singleRxAnt} will be used in deriving our statistical analysis, and that our RRH placement problem will depend on these derived average expressions, thus the analysis is a statistical one and will not depend on the instantaneous channels.

\subsubsection{Multicast for RRHs with multiple-receive antennas}
Each RRH uses maximum ratio combining (MRC) to combine its $M$ received signals~\cite{goldsmith2005wireless, blockdiagonalization4907459}. To account for the multiple receive antennas, the CU uses the asymptotic multicast beamformer
\begin{align}\label{eq:asymptoticBeamformer_MRxAnt}
	{\bf \bar{w}}_{q} = \bar{\mu}_{q} \sum_{n=1}^{N} \frac{{\bf \bar{H}}_{q,qn} {\bf b}_{qn}}{\bar{\ell}_q(x_{qn},y_{qn})} .
\end{align}
Here, ${\bf \bar{H}}_{q,qn}^*$ is the $M_{\rm c} \times M$ channel between CU $q$ and RRH $n$ in cell $q$, and ${\bf b}_{qn} \in \mathbb{C}^{M}$ is the combining weight vector at RRH $n$. To implement MRC, we use the Singular Value Decomposition (SVD) of ${\bf \bar{H}}_{q,qn} = {\bf U}_{q,qn} \bm{\Sigma}_{q,qn} {\bf V}_{q,qn}^H$ where ${\bf V}_{q,qn} = [{\bf v}_{q,qn1}\ {\bf v}_{q,qn2}\ \dots\ {\bf v}_{q,qnM}]$, and the required combining weight vector is ${\bf b}_{qn} = {\bf v}_{q,qn1}$, i.e., \emph{the singular vector corresponding to the largest singular value}. As before, the term $\bar{\mu}_{q}$ is a common factor to normalize ${\bf \bar{w}}_{q}$, i.e., $\mathbb{E}\left\{\|{\bf \bar{w}}_{q}\|^2\right\}=1$, but is now based on $\mathbb{E}\left\{\sigma_{q,qn}^{(G)}\right\}$, the mean of the \emph{largest singular value} of the $M_{\rm c} \times M$ Gaussian small-scale fading matrix ${\bf \bar{G}}_{q,qn}$ (we note that this mean is independent of the RRH). Specifically, we have
\begin{align} \label{eq:mu2_formula}
	\bar{\mu}_{q} = \frac{1}{ \mathbb{E}\left\{\sigma_{q,qn}^{(G)}\right\} \sqrt{\sum_{n=1}^{N} \frac{1}{\bar{\ell}_q(x_{qn},y_{qn})}}} 
\end{align}

Surprisingly, there is no closed-form expression for the mean of the largest singular value of a Gaussian random matrix. The work in~\cite{rudelson2010non} indicates that this mean is \emph{lower than} $\sqrt{M}+\sqrt{M_{\rm c}}$. In our testing, we found the approximation $\mathbb{E}\big\{\sigma_{q,qn}^{(G)}\big\} \simeq \big(\sqrt{M/2} + \sqrt{M_{\rm c}}\big)$ to be the accurate.

\subsection{Statistical Analysis of the Multicast Fronthaul Channel}
Due to fading, the multicast fronthaul rate is random and hence, we can only place a probabilistic outage constraint on the fronthaul. This, in turn, requires a statistical analysis of the signals received on the fronthaul. Specifically, we impose the constraint 
\begin{equation}
	\mathbb{P}\left\{\min_{n} R_{qn}^{\rm{(b)}}\left({\bf x}_{q},{\bf y}_{q}\right) \le 
	K \frac{\omega}{\omega_{\rm b}} 
	\mathbb{E}_{\widetilde{x}_k,\widetilde{y}_k}\left\{ R_{qk}^{\rm{(a)}}({\bf x},{\bf y}) \right\}
	\right\}\le\epsilon,
	\label{eq:Opt2Constraint1}
\end{equation}
where $\omega$, $\omega_{\rm b}$ are the bandwidth allocated for the data plane for the access channel and fronthaul respectively, and $\epsilon$ denotes the allowed fronthaul outage probability. The constraint sets the probability of outage to be less than a chosen value of $\epsilon$.
Here, $R_{qn}^{\rm{(b)}}\left({\bf x}_{q},{\bf y}_{q}\right) = \log\big(1 + \frac{S_{qn}^{(b)}}{I_{qn}^{(b)}+\sigma_z^2}\big)$ denotes the achievable rate on the fronthaul. We must work with the minimum rate in the fronthaul because all $N$ RRHs must receive the data for the users. The right hand side of the inequality denotes the average rate to the $K$ users, with 
\begin{align} \label{eq:trafficExpectation}
	\mathbb{E}_{\widetilde{x}_k,\widetilde{y}_k}\left\{ R_{qk}^{\rm{(a)}}({\bf x},{\bf y}) \right\} = 
	\int \!\!  \int_{\widetilde{x}_k,\widetilde{y}_k\in \mathcal{B}_{q}} R_{qk}^{\rm{(a)}}({\bf x},{\bf y}) f_q(\widetilde{x}_k,\widetilde{y}_k) \diff \widetilde{x}_k \diff \widetilde{y}_k
\end{align}
where $\mathcal{B}_{q}$ is the boundary of cell $q$. The probabilistic outage constraint for the fronthaul in~\eqref{eq:trafficExpectation} cannot be directly used to optimize the location of the RRHs, hence some simplification is needed. A natural approach is to use some reference distribution to characterize this constraint. To do this, we fit the signals received at the RRHs on the fronthaul with a reference distribution, then we use this framework to characterize the constraint in~\eqref{eq:trafficExpectation}.

\begin{figure*}[t]
	\centering
	\begin{subfigure}[t]{0.31\textwidth}
		\centering
		\includegraphics[width=0.95\textwidth]{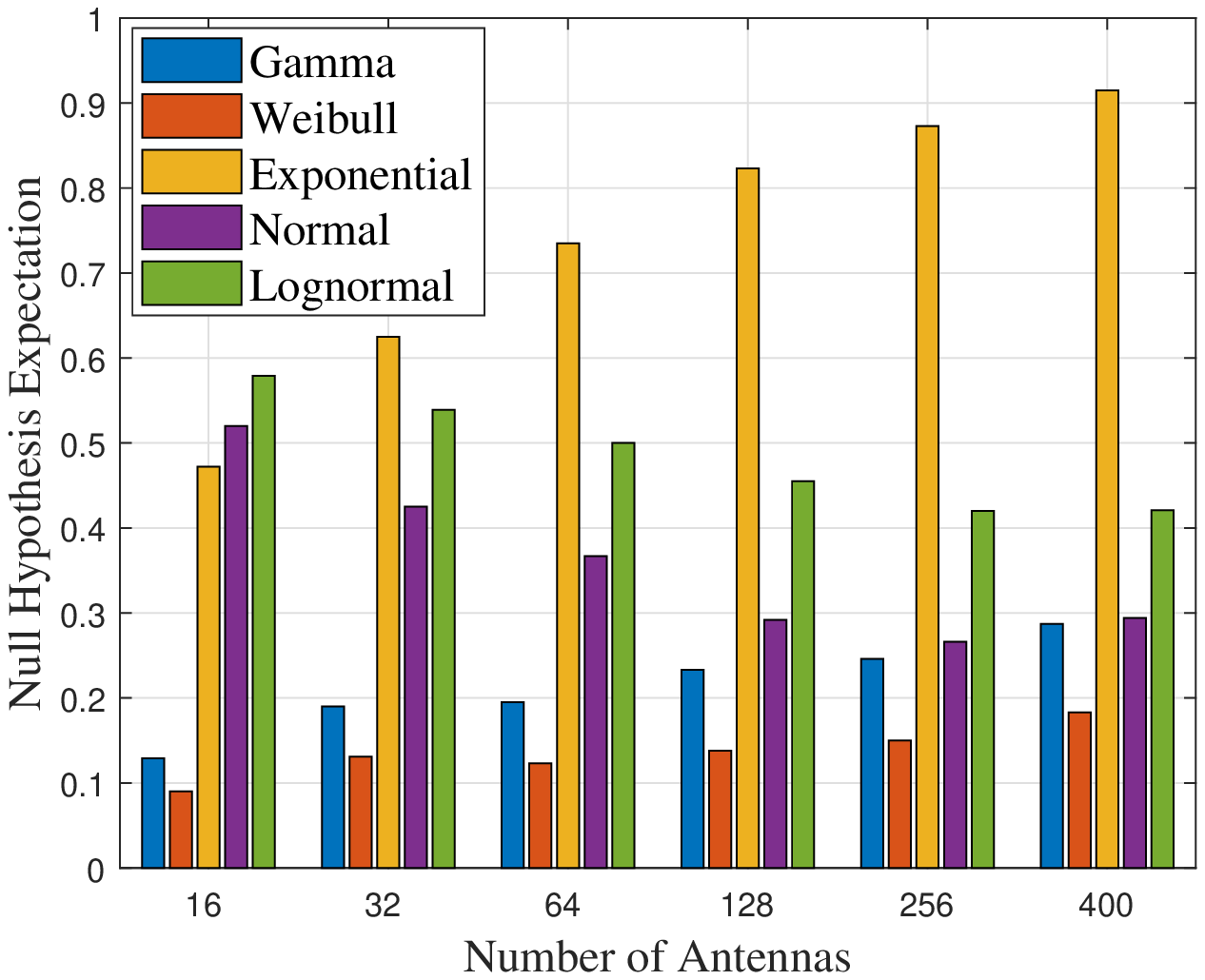}
		\caption{Desired signal power using multicast with multiple receive antennas.}
		\label{fig:SignalPower_Multicast_M8_PaperModel}
	\end{subfigure}
	$\ $
	\begin{subfigure}[t]{0.31\textwidth}
		\centering
		\includegraphics[width=0.95\textwidth]{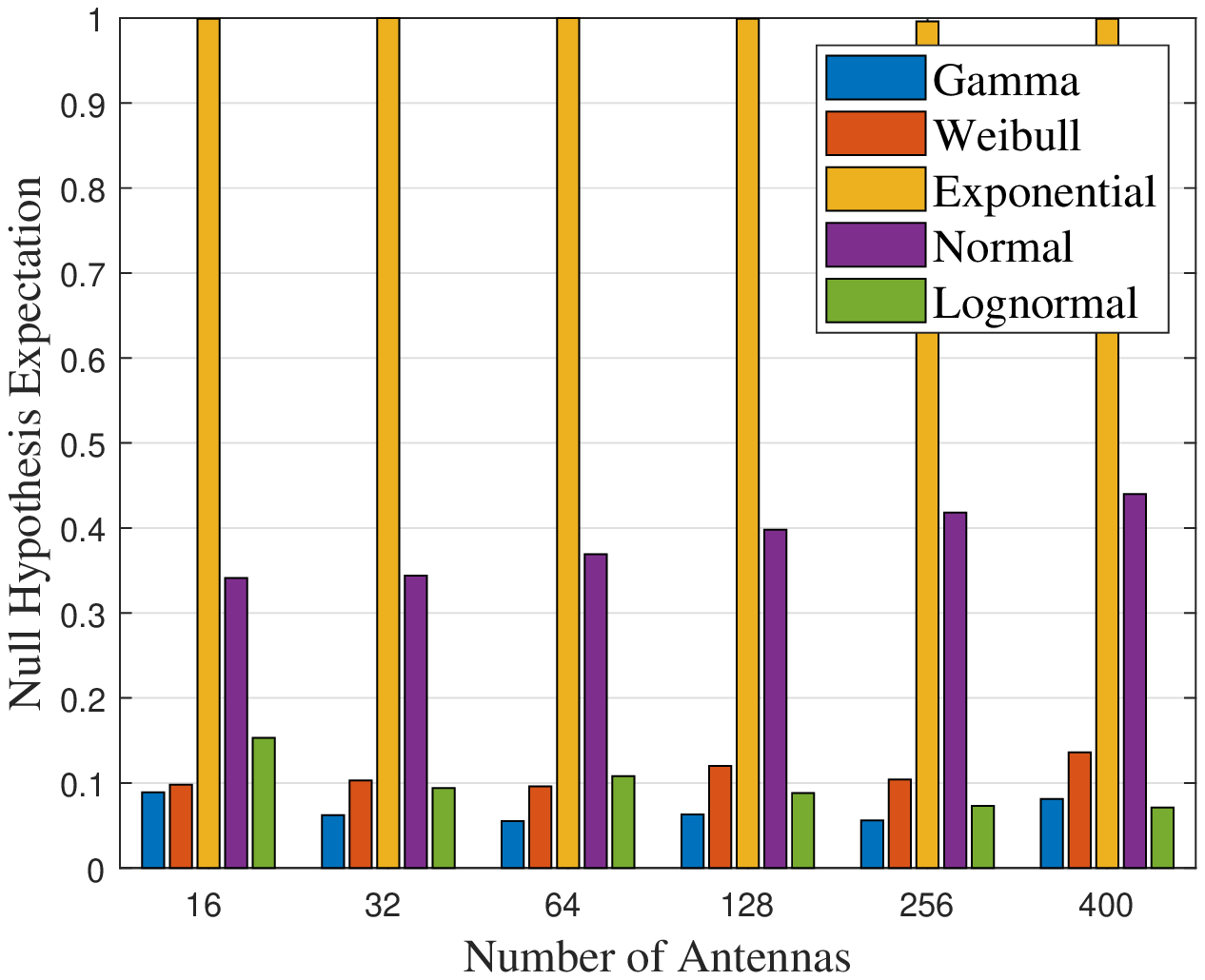}
		\caption{Interference power using multicast with multiple receive antennas.}
		\label{fig:InterferencePower_Multicast_M8_PaperModel}
	\end{subfigure}
	$\ $
	\begin{subfigure}[t]{0.31\textwidth}
		\centering
		\includegraphics[width=0.95\textwidth]{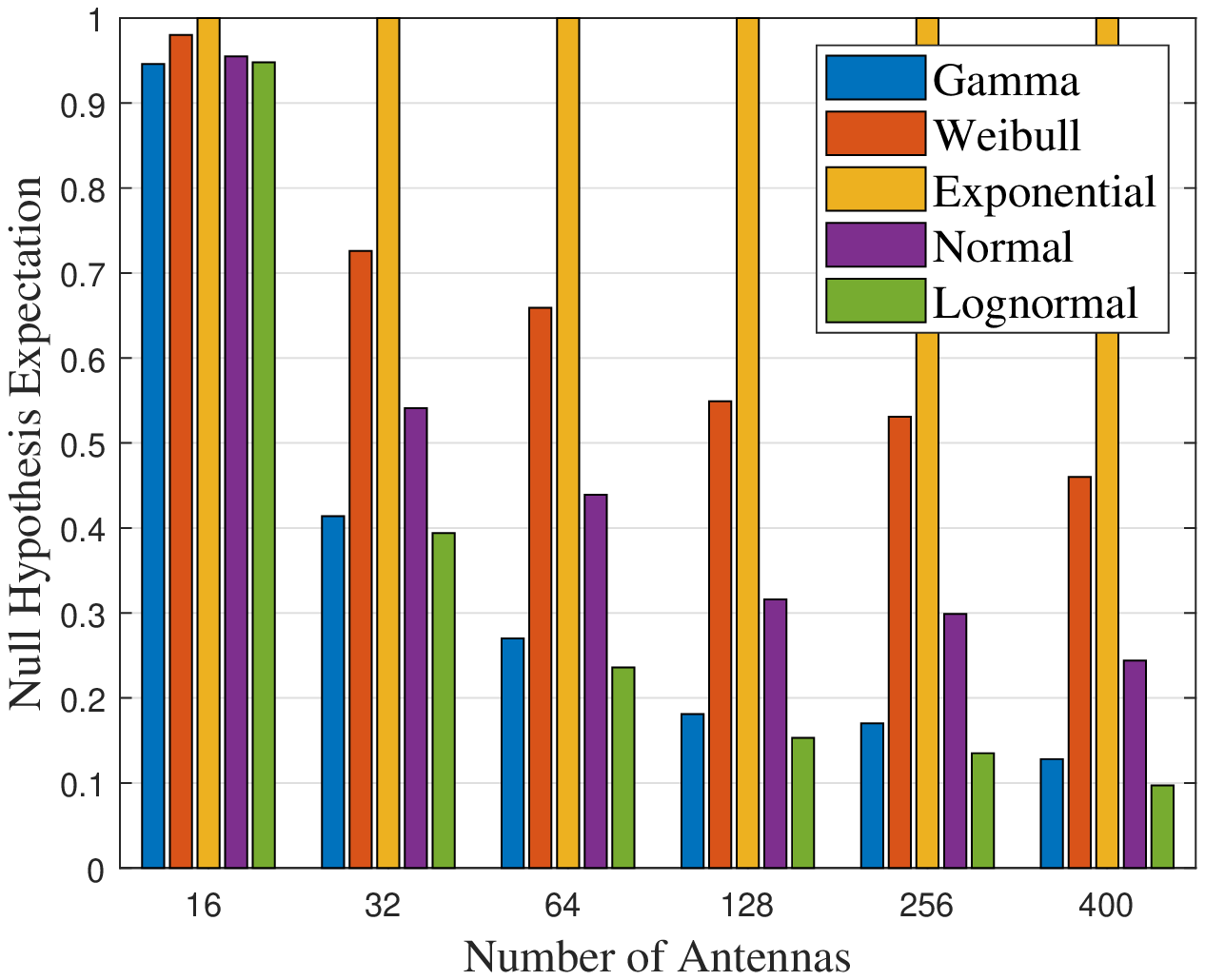}
		\caption{Interference power using ZF with multiple receive antennas (desired signal power is constant).}
		\label{fig:InterferencePower_ZF_M8_PaperModel}
	\end{subfigure}
	\vspace{-0.5em}
	\caption{Result of KS test performed on samples from Monte Carlo simulation; distribution are fit using our derived expressions for mean and variance.}
	\label{fig:KSTest}
	\vspace{-1em}
\end{figure*}

We apply the KS test on the signals received at the RRHs on the fronthaul. This test compares a sample statistic with a reference distribution to reject (null hypothesis) or accept the hypothesis that the sample was drawn from the reference distribution. Hence, it estimates the goodness of the fit. In Fig.~\ref{fig:KSTest}, we plot the average number of times the null hypothesis was retained.

For the test setup, we use $Q=9$ square cells with wraparound (to eliminate network boundary effects), where the CU for each cell is found at the cell center, i.e., the CUs are located on a grid. Each cell contains $N=10$ RRHs randomly deployed in the cell with an exclusion region of $20$ meters around the CU. Other details on the network setup can be found in Table~\ref{table:sim_parameters}. We generate Monte Carlo simulations for $100$ network realizations, each has $1000$ channels realizations. We then apply the KS test to each set of channel realizations using many reference distributions, shown in Fig.~\ref{fig:KSTest}. Finally, we obtain the average number of times that the null hypothesis was retained in, i.e., we obtain the average number of times in which the hypothesis that the samples came from the reference distribution was rejected.

Our results show that the gamma distribution has the lowest null hypothesis tests obtained from our Monte Carlo simulation. Thus, the gamma distribution is the best fit for the different beamforming techniques. Based on this, we will derive accurate closed-form expressions for the mean and variance of the power of the desired signal and interference. Then, we will relate these moments to the parameters of the gamma distribution to analytically perform the fitting.

Based on this discussion, we model the desired signal and interference powers at each RRH as a gamma RV $Y$, which is characterized by shape ($K_Y$) and scale ($\theta_Y$) parameters where
\begin{align} \label{eq:shape_scale_GammaDist}
K_Y = \frac{\left(\mathbb{E}\{Y\}\right)^2}{\mathtt{Var}\{Y\}} > 0, \hspace*{0.2in} \theta_Y = \frac{\mathtt{Var}\{Y\}}{\mathbb{E}\{Y\}} > 0
\end{align}
We therefore need the mean and variance of the signal and interference powers to characterize the gamma distribution. However, unlike other studies using a similar model that use numerical methods~\cite{8449213, 7898403}, we provide accurate, if approximate, closed-form expressions for these moments.

\noindent \textit{Multicast Fronthaul for RRHs with single-receive antennas:} The derivation of the required closed-form expressions is quite involved, and we defer the details to an appendix. Specifically, the appendix shows, \textit{for a single receive antenna} using the asymptotic beamformer defined in~\eqref{eq:asymptoticBeamformer_singleRxAnt}, we can closely approximate the mean and the variance of $S_{qn}^{(b)}$, the desired signal power received at RRH $n$ from CU $q$ in the multicast case as
	{\allowdisplaybreaks
	\begin{align}
	\mathbb{E}\left\{S_{qn}^{(b)}\right\} & \simeq
	p_{\rm c} \bar{\mu}_q^2 \bar{\ell}_q(x_{qn},y_{qn})\left( \frac{M_{\rm c}^2}{\bar{\ell}_q(x_{qn},y_{qn})}
	+ M_{\rm c} \sum_{n' \ne n}^{N} \frac{1}{\bar{\ell}_q(x_{qn'},y_{qn'})}
	\right)
	\label{eq:SigfronthaulMean_Multicast}
	\\
	\mathtt{Var}\left\{S_{qn}^{(b)}\right\} & \simeq \left(p_{\rm c} \bar{\mu}_q^2 \bar{\ell}_q(x_{qn},y_{qn}) \right)^2
	\left( 4M_{\rm c}^3 \left(\frac{1}{\bar{\ell}_q(x_{qn},y_{qn})}\right)^2
	+ \frac{2 M_{\rm c}^3}{\bar{\ell}_q(x_{qn},y_{qn})}\sum_{n' \ne n}^{N} \frac{1}{\bar{\ell}_q(x_{qn'},y_{qn'})}
	\right.
	\nonumber \\
	&\quad
	\quad \quad \quad \quad
	\left.
	+ M_{\rm c}^2 \sum_{n' \ne n}^{N} \frac{1}{\bar{\ell}_q(x_{qn'},y_{qn'})} \sum_{n'' \ne n}^{N} \frac{1}{\bar{\ell}_q(x_{qn''},y_{qn''})}
	\right) \label{eq:SigfronthaulVar_Multicast}
	\end{align}
	}
	For the same scenario, the mean and the variance of the interference power $I_{qn}^{(b)}$ is
	{\allowdisplaybreaks
	\begin{align}
	\mathbb{E}\left\{I_{qn}^{(b)} \right\} & \simeq p_{\rm c} \sum_{q' \ne q}^{Q} \bar{\mu}_{q'}^2 \bar{\ell}_{q'}(x_{qn},y_{qn}) M_{\rm c} \sum_{n' = 1}^{N} \frac{1}{\bar{\ell}_{q'}(x_{q'n'},y_{q'n'})}
	= p_{\rm c} \sum_{q' \ne q}^{Q} \bar{\ell}_{q'}(x_{qn},y_{qn})  \label{eq:IntfronthaulMean_Multicast}
	\\
	\mathtt{Var}\left\{I_{qn}^{(b)}\right\}
	& \simeq
	\sum_{q' \ne q}^{Q} \left(p_{\rm c} \bar{\mu}_{q'}^2 \bar{\ell}_{q'}(x_{qn},y_{qn}) M_{\rm c} \sum_{n' = 1}^{N} \frac{1}{\bar{\ell}_{q'}(x_{q'n'},y_{q'n'})} \right)^2
	= \sum_{q' \ne q}^{Q} \left(p_{\rm c} \bar{\ell}_{q'}(x_{qn},y_{qn}) \right)^2 
	\label{eq:IntfronthaulVar_Multicast}
	\end{align}}
We refer the reader to Appendix~\ref{appendix:meanVar_singleRXOutage} for the detailed derivations. As our results show, these approximations are extremely accurate.

\noindent \textit{Multicast Fronthaul for RRHs with multiple-receive antennas:} We next generalize the expressions in~\eqref{eq:SigfronthaulMean_Multicast}-\eqref{eq:IntfronthaulVar_Multicast} to the case where the RRH uses all its $M$ antennas for reception on the fronthaul.

The received SINR at the fronthaul of RRH $n$ is
\begin{align}
\text{SINR}_{qn}^{(b)} = \frac{S_{qn}^{(b)}}{I_{qn}^{(b)}+\sigma_z^2}=
\frac{p_{\rm c}
	|{\bf b}_{qn}^H {\bf \bar{H}}_{q,qn}^H {\bf \bar{w}}_{q}|^2
}
{
	p_{\rm c} \sum_{q'\ne q}^{Q} |{\bf b}_{qn}^H {\bf \bar{H}}_{q',qn}^H {\bf \bar{w}}_{q'}|^2
	+ \sigma_z^2
}.
\end{align}

By using the SVD and the combining vector, we can extend the analysis for the single antenna receiver to the case of $M$ receive antennas. Appendix~\ref{appendix:multicast_multiRX_outage} details this extension. Specifically, assuming $M$ receive-antennas per RRH, and using the asymptotic beamformer defined in~\eqref{eq:asymptoticBeamformer_MRxAnt}, we can closely approximate the mean and the variance of $S_{qn}^{(b)}$, the desired signal power received at RRH $n$ from CU $q$ in the multicast case as
	{\allowdisplaybreaks
	\begin{align}
	\mathbb{E}\left\{S_{qn}^{(b)}\right\} & \simeq p_{\rm c} \bar{\mu}_q^2 \bar{\ell}_q(x_{qn},y_{qn}) \left(\mathbb{E}\left\{\sigma_{q,qn}^{(G)}\right\}\right)^4 
	\bigg( \frac{1}{\bar{\ell}_q(x_{qn},y_{qn})} + \frac{1}{M_{\rm c}^2\sqrt{\bar{\ell}_q(x_{qn},y_{qn})}} \sum_{n' \ne n}^{N} \frac{1}{\bar{\ell}_q(x_{qn'},y_{qn'})}
	\nonumber \\[-10pt]
	&
	\quad \quad \quad
	+ \sum_{n' \ne n}^{N} \bigg( \frac{1}{M_{\rm c} \bar{\ell}_q(x_{qn'},y_{qn'})}
	+ \frac{1}{M_{\rm c}^2 \sqrt{\bar{\ell}_q(x_{qn},y_{qn})\bar{\ell}_q(x_{qn'},y_{qn'})}} \bigg)
	\bigg) \label{eq:MultAntSigMean}
	\\[-5pt]
	\mathtt{Var}\left\{S_{qn}^{(b)}\right\} & \simeq \left(p_{\rm c} \bar{\mu}_q^2 \bar{\ell}_q(x_{qn},y_{qn})\right)^2 \left(\mathbb{E}\left\{\sigma_{q,qn}^{(G)}\right\}\right)^8 
	\bigg(
	\frac{2}{M_{\rm c} \bar{\ell}_q(x_{qn},y_{qn})}\sum_{n' \ne n}^{N} \frac{1}{\bar{\ell}_q(x_{qn'},y_{qn'})}
	\nonumber \\[-10pt]
	& \quad \quad \quad \quad \quad
	+ \frac{1}{M_{\rm c}^2} \sum_{n' \ne n}^{N} \frac{1}{\bar{\ell}_q(x_{qn'},y_{qn'})} \sum_{n'' \ne n}^{N} \frac{1}{\bar{\ell}_q(x_{qn''},y_{qn''})}
	\bigg) \label{eq:MultAntSigVar}
	\end{align}
	Similarly for $I_{qn}^{(b)}$, the interference power received at RRH $n \in \mathcal{D}_q$ is
	\vspace{-0.3em}
	\begin{align}
	\mathbb{E}\left\{I_{qn}^{(b)} \right\} &= 
	p_{\rm c} \sum_{q' \ne q}^{Q} \bar{\mu}_{q'}^2 \bar{\ell}_{q'}(x_{qn},y_{qn}) \left(\mathbb{E}\left\{\sigma_{q,qn}^{(G)}\right\}\right)^2 \sum_{n' = 1}^{N} \frac{1}{\bar{\ell}_{q'}(x_{q'n'},y_{q'n'})}
	= p_{\rm c} \sum_{q' \ne q}^{Q} \bar{\ell}_{q'}(x_{qn},y_{qn}) \label{eq:MultAntIntMean}
	\end{align}
	\vspace{-0.5em}
	\begin{align}
	\mathtt{Var}\left\{I_{qn}^{(b)}\right\}
	&= \sum_{q' \ne q}^{Q} \left(p_{\rm c} \bar{\mu}_{q'}^2 \bar{\ell}_{q'}(x_{qn},y_{qn}) \right)^2 \left(\mathbb{E}\left\{\sigma_{q,qn}^{(G)}\right\}\right)^4
	\sum_{n' = 1}^{N} \frac{1}{\bar{\ell}_{q'}(x_{q'n'},y_{q'n'})} \sum_{n'' = 1}^{N} \frac{1}{\bar{\ell}_{q'}(x_{q'n''},y_{q'n''})}
	\nonumber \\[-6pt]
	&= \sum_{q' \ne q}^{Q} \left(p_{\rm c} \bar{\ell}_{q'}(x_{qn},y_{qn}) \right)^2 \label{eq:MultAntIntVar}
	\end{align}
	}
\noindent \textit{Statistical Outage Analysis}: Having provided closed-form expressions for the mean and variance of the desired signal and interference powers, we use~\eqref{eq:shape_scale_GammaDist} to obtain the required corresponding scale and shape parameters, thereby specifying the gamma RVs for these powers.

To obtain a closed-form expression for the constraint in~\eqref{eq:Opt2Constraint1}, we use the fact that $R_{qn}^{\rm{(b)}} < \dot{\mathsf{x}}$ if $S^{\rm{(b)}}_{qn} < (e^{\dot{\mathsf{x}}}-1)(I^{\rm{(b)}}_{qn} + \sigma_z^2)$ i.e., $\mathbb{P}\left\{ R_{qn}^{\rm{(b)}} < \dot{\mathsf{x}} \mid I^{\rm{(b)}}_{qn}\right\} = F_{S^{\rm{(b)}}_{qn}}\left((e^{\dot{\mathsf{x}}}-1)(I_{qn} + \sigma_z^2)\right)$ where $F_{S^{\rm{(b)}}_{qn}}(\cdot)$ denotes the cumulative density function (CDF) of the signal power (gamma) RV $S^{\rm{(b)}}_{qn}$. In turn, averaging over the interference power, we have
\begin{align}\label{eq:CDF}
F_{R_{qn}^{\rm{(b)}}}(\dot{\mathsf{x}}) = \int_{0}^{\infty}
F_{S^{\rm{(b)}}_{qn}}\left(\left(e^{\dot{\mathsf{x}}} - 1\right) \left(i + \sigma_z^2\right)\right) f_{I^{\rm{(b)}}_{qn}}\left(i\right)
\diff i,
\end{align}
where $f_{I^{\rm{(b)}}_{qn}}(\cdot)$ denotes the PDF of the interference, $I^{\rm{(b)}}_{qn}$ (also approximated as a gamma distribution). Finally, by defining $\bar{R}_q = \min_n R_{qn}^{\rm{(b)}}\left({\bf x}_{q},{\bf y}_{q}\right)$, we have 
\vspace{-0.2em}
\begin{align}\label{eq:multicastfronthaulOutage}
	F_{\bar{R}_q}(\dot{\mathsf{x}}) = 1 - \prod_{n=1}^{N} \left[1 - F_{R_{qn}^{\rm{(b)}}}(\dot{\mathsf{x}})\right],
\end{align}
where, in the constraint of~\eqref{eq:Opt2Constraint1}, $\dot{\mathsf{x}} = K \frac{\omega}{\omega_{\rm b}} \mathbb{E}_{\widetilde{x}_k,\widetilde{y}_k}\big\{ R_{qk}^{\rm{(a)}}({\bf x},{\bf y}) \big\}$. In the next section, we will optimize the placement of RRHs with a statistical outage constraint on the fronthaul rate.

\subsection{RRHs Placement Under Multicast Fronthaul} \label{subsec:PlacementMulticastAlg}
In the previous section we provided a statistical analysis of the fronthaul. We are now ready to formulate our location optimization problem, with the RRH locations as the optimization variables. In theory, the algorithm should optimize the locations of the $QN$ RRHs in all $Q$ cells jointly. However, we found this problem to be intractable and, hence, we exploit the sub-optimal, but well-accepted, approach of block coordinate descent. Specifically, we optimize the $N$ RRH locations in a cell (say $q$, i.e., $({\bf x}_q, {\bf y}_q)$), while keeping the RRH locations in the other $(Q-1)$ cells (i.e., $({\bf x}_{-q}, {\bf y}_{-q})$) fixed; we then iterate over all cells until convergence. 

Within the cell being optimized, our objective function is the average access rate (rate for a typical user) as given in~\eqref{eq:trafficExpectation}. We maximize this rate with the outage constraint as in~\eqref{eq:Opt2Constraint1}, rewritten in~\eqref{eq:multicastfronthaulOutage}. Therefore, for cell $q$, our optimization problem is
%
\begin{align}\label{eq:multicast_singleRxOpt}
	&\underset{ {\bf x}_q, {\bf y}_q}{\max}\quad
	\mathbb{E}_{\widetilde{x}_k,\widetilde{y}_k}\left\{ R_{qk}^{\rm{(a)}}({\bf x},{\bf y}) \right\}	\hspace*{0.3in} \text{s.t.} \hspace*{0.3in} F_{\bar{R}_q}(\dot{\mathsf{x}}) \leq \epsilon 
\end{align}
where $\dot{\mathsf{x}}$, as defined below~\eqref{eq:multicastfronthaulOutage}, also includes the optimization variables. As shown in~\eqref{eq:trafficExpectation}, the mean is over the locations of the users in the cell. The constraint in~\eqref{eq:multicast_singleRxOpt} restricts the fronthaul outage to a maximum value $\epsilon$, i.e., the RRH locations are forced to be close enough to the CU such that the fronthaul can sustain the mean access channel capacity with probability $(1-\epsilon)$. 

We solve this optimization problem using a gradient-based method. The derivative of the objective function in~\eqref{eq:multicast_singleRxOpt} with respect to the $x$-coordinate $x_{qm}$ of RRH $m\in \mathcal{D}_q$ is given by
\begin{align}\label{eq:locationsUpdate}
\nabla_{x_{qm}} &= \frac{\partial \mathbb{E}_{\widetilde{x}_k,\widetilde{y}_k}\big\{ R_{qk}^{\rm{(a)}}({\bf x},{\bf y}) \big\}}{\partial x_{qm}}
= 
\frac{\partial \int \!\!  \int_{\widetilde{x}_k,\widetilde{y}_k\in \mathcal{B}_q} \!\! R_{qk}^{\rm{(a)}}({\bf x},{\bf y}) f_q(\widetilde{x}_k,\widetilde{y}_k)
	\diff \widetilde{x}_k \diff \widetilde{y}_k }{\partial x_{qm}}
\nonumber \\[-5pt]
&=\ 
\int \!\!  \int_{\widetilde{x}_k,\widetilde{y}_k\in \mathcal{B}_q} \!\!
\Theta_m(\widetilde{x}_k,\widetilde{y}_k)
f_q(\widetilde{x}_k,\widetilde{y}_k)
\diff \widetilde{x}_k \diff \widetilde{y}_k,
\end{align}
with
\begin{align} \label{eq:Theta_m}
\Theta_m(\widetilde{x}_k,\widetilde{y}_k)
= \frac{ \left(\widetilde{x}_k - x_{qm}\right) \left(\gamma_k\left({\bf x}_{-q},{\bf y}_{-q}\right)\right)^{-1} \left(1+d_{qmk}/d_0\right)^{-1-\alpha} }{ d_{qmk}\big(1+\gamma_k\left({\bf x}_{-q},{\bf y}_{-q}\right)^{-1}
	\sum_{n\in\mathcal{D}_q}\left(1+d_{qnk}/d_0\right)^{-\alpha}\big)},
\end{align}
A similar expression can be obtained for the gradient with respect to the $y$-coordinate. We note that the expression in~\eqref{eq:Theta_m} depends also on the locations $\left({\bf x}_{-q},{\bf y}_{-q}\right)$, which as noted earlier, represent the locations of the RRHs in all cells other than cell $q$.

The \emph{overall optimization problem} across all $Q$ cells can be solved using Algorithm~\ref{algorithm:iterativeAlgoMulticast}. The problem in~\eqref{eq:multicast_singleRxOpt} represents the inner loop (steps~\ref{step:iterativeAlgoMulticast_FlagCheckStart}-\ref{step:iterativeAlgoMulticast_d_q}), where we update the locations of the RRHs in cell $q$ once using the gradient direction and a step-size $\nu$ (which is scaled to ensure any step taken retains the feasibility of RRH locations). Specifically, on iteration $i$
\begin{align}\label{eq:multicast_xmymUpdate}
	x_{qm}^{(i+1)} = x_{qm}^{(i)} + \nu \nabla_{x_{qm}},
	\hspace*{0.3in} y_{qm}^{(i+1)} = y_{qm}^{(i)} + \nu \nabla_{y_{qm}}
\end{align}
		
As can be seen in the first step in the algorithm (Step~\ref{step:iterativeAlgoMulticast_1}), we start by generating initial feasible locations for the RRHs, for example inside a small region around their serving CUs, which respect the fronthaul constraint. If the initial locations for the RRHs, which are generated inside a small region around their serving CUs (Step~\ref{step:iterativeAlgoMulticast_1} in Algorithm~\ref{algorithm:iterativeAlgoMulticast}), satisfy the constraint for the fronthaul outage, then we already have a feasible solution, and the algorithm can either retain the current solution or obtain a better one. On the other hand, if the initial locations close to the CU do not satisfy the constraint, then there is no feasible solution, and a more relaxed fronthaul constraint is needed (e.g., by decreasing the number of served users $K$ or allocating more bandwidth for the fronthaul). If the initial locations do not satisfy the constraint then there is no point in optimizing the locations of the RRHs because as the RRHs move away from the CU, the fronthaul constraint will become tighter.

\setlength{\textfloatsep}{0pt}
\begin{algorithm}[t]
	\SetAlgoLined
	\SetInd{0.1em}{1em}
	\caption{RRHs Location Optimization for Multicast}
	\label{algorithm:iterativeAlgoMulticast}
	\footnotesize
	\textbf{Input:} Location of CUs, $Q, N, K, M$, PDF of spatial traffic distribution $\{f_q(\widetilde{x}_k,\widetilde{y}_k) : q = 1, \dots, Q\}$, bandwidths ($\omega$, $\omega_b$), fronthaul constraint, $\epsilon$, convergence criterion $d_{\rm cvg}$, boundaries of cells $\mathcal{B}_q$. \\
	\textbf{Outputs:} Locations of RRHs $\{(\mathbf{x}_q, \mathbf{y}_q): q = 1, \dots, Q\}$.\\
	Generate feasible locations for the RRHs in each cell $q$ and choose $d_\text{max} > d_\text{cvg}$. \label{step:iterativeAlgoMulticast_1}\\
	\While{$d_\text{max} > d_\text{cvg}$}{\label{step:iterativeAlgoMulticast_stepSizeUpdate}
		\For{$q \in \{1,\dots,Q\}$}{
			Wraparound cells to make cell $q$ at center. \label{step:iterativeAlgoMulticast_wrap}\\
			Flag = 1\\
			\While{Flag = 1}{\label{step:iterativeAlgoMulticast_FlagCheckStart}
				\For{$m \in \mathcal{D}_q$}{
					Obtain $x_{qm}^{\rm temp}$ and $y_{qm}^{\rm temp}$ using \eqref{eq:multicast_xmymUpdate} \label{step:x_y_temp}.   \\
				}
				\eIf{constraint in \eqref{eq:multicast_singleRxOpt} using $({\bf x}_{q}^{\rm temp}$,${\bf y}_{q}^{\rm temp})$ not satisfied \textbf{or} $({\bf x}_{q}^{\rm temp}$,${\bf y}_{q}^{\rm temp})$ not within cell $q$
				}{\label{step:iterativeAlgoMulticast_constraintCheck}
					$\nu = \nu_{f}\nu $ \label{step:shrinkStepSize}\\
				}{
					Update ${\bf x}_{q}^{(i+1)} = {\bf x}_{q}^{\rm temp}$ and ${\bf y}_{q}^{(i+1)} = {\bf y}_{qm}^{\rm temp}$ \label{step:locationUpdate}\\
					Flag = 0 
				}
			}\label{step:iterativeAlgoMulticast_FlagCheckEnd}
			$d_q=\underset{n}{\max}\left\{\underset{x,y}{\max}\left\{|x_{qn}^{(i+1)} - x_{qn}^{(i)}|, |y_{qn}^{(i+1)} - y_{qn}^{(i)}|\right\}\right\}$ \label{step:iterativeAlgoMulticast_d_q}
		}
		$d_\text{max} = \underset{q}\max\ d_q$
	}
	\vspace{-0.5em}
\end{algorithm}

Steps~\ref{step:iterativeAlgoMulticast_FlagCheckStart}-\ref{step:iterativeAlgoMulticast_d_q} inside the \texttt{while} loop update the $x$ and $y$ coordinates of the RRHs in cell $q$ by using~\eqref{eq:multicast_xmymUpdate}. The algorithm refers to the new coordinates as temporary (Step~\ref{step:x_y_temp}) to confirm that these coordinates satisfy the fronthaul constraint and the RRH remains within the cell (Step~\ref{step:iterativeAlgoMulticast_constraintCheck}). Here $\nu$ represents the step size, which, if the constraint is not satisfied, is reduced by a factor of $\nu_f < 1$ (as shown in Step~\ref{step:shrinkStepSize}). We note that keeping the RRHs within their cell can be easily achieved by setting the PDF of the spatial traffic distribution to zero outside the cell boundary. In this case, the locations of the RRHs will move following the gradient direction~\eqref{eq:multicast_xmymUpdate}, and by setting the PDF of the traffic distribution to zero outside the cell boundary, we can eliminate the possibility of having a gradient direction pointing outside the cell. Thus, the x- and y-coordinates of the RRHs will not be steered toward the outside boundary of the cell.

Step~\ref{step:locationUpdate} indicates that the algorithm executes one iteration for each cell; thus the total number of iterations for all cells is the same, and each iteration uses the updated RRH locations in the interfering cells until convergence, where the maximum change in the locations of the RRHs $d_\text{max}$ becomes less than a convergence criterion of $d_\text{cvg}$. With a proper value for $d_\text{cvg}$, the algorithm is guaranteed to converge to a stationary point; in our testing $d_\text{cvg} = 1$~m is an appropriate value. In our section of numerical results, we illustrate the convergence of the proposed algorithm.

To analyze the computational complexity of our algorithm, we have a complexity of $\mathcal{O}\left( Q N \Omega_1 \right)$ to update the temporary locations of the RRHs using~\eqref{eq:multicast_xmymUpdate}, as shown in Step~\ref{step:x_y_temp}, where $\Omega_1$ is the complexity of performing the numerical double integration in~\eqref{eq:locationsUpdate} over the region of the cell $\mathcal{B}_q$ which is chosen as a square. We also have a complexity of repeating Step~\ref{step:x_y_temp} using a shrinking step size (Step~\ref{step:shrinkStepSize}) for the gradient descent in case the fronthaul constraint is not satisfied, leading to $\zeta$ repetitions. For the calculation of the fronthaul outage, we have a complexity of $\mathcal{O}\left( Q N \right)$ to update the values of the moments of the power of the useful and interference signals, another $\mathcal{O}\left( Q N \right)$ to update the parameters of the gamma distribution, and finally a complexity of $\mathcal{O}\left( Q N \Omega_2 \right)$ to perform the numerical integration needed to evaluate the fronthaul outage in~\eqref{eq:CDF} for the $Q$ cells, where $\Omega_2$ is the complexity of performing the numerical single integration in~\eqref{eq:CDF}. Thus, we have a total algorithm complexity of $\mathcal{O}\left( \zeta Q N \Omega_1 + Q N \Omega_2 \right)$ per iteration. Our results shown in Section~\ref{sec:results} indicate that the algorithm converges in $10$-$15$ iterations.

\section{Fronthaul Using Zero Forcing Beamforming}\label{sec:fronthaul_ZFBF}
In the previous section, we analyzed the use of multicast in the fronthaul. However, as we will see, by communicating to all $N$ RRHs using the same beamformer, the achievable rate for large $N$ is low. In this section, we analyze an alternative scheme, where each CU uses ZF beamforming on the fronthaul to communicate to the $N$ RRHs in its cell. Using ZF eliminates intra-cell interference but suffers from a 
power penalty since the CU must communicate to each RRH independently.

As in the previous section, each RRH with $M$ antennas uses MRC to provide receive diversity. Specifically, as in~\eqref{eq:asymptoticBeamformer_MRxAnt}, this is done through the combining vector ${\bf b}_{qn} \in \mathbb{C}^M$, the right singular vector corresponding to the largest singular value of ${\bf \bar{H}}_{q,qn} \in \mathbb{C}^{M_{\rm c} \times M}$. In turn, the ZF beamformer is constructed in each cell $q$ separately as
\begin{align}\label{eq:beamformerZFfronthaul}
{\bf \bar{W}}_{q}^{\rm{(b)}}={\bf \widetilde{W}}_{q}^{\rm{(b)}} \bm{\bar{\mu}}_q \qquad \text{where} \qquad {\bf \widetilde{W}}_{q}^{\rm{(b)}}
= {\bf E}_{q}\left({\bf E}_{q}^H{\bf E}_{q}\right)^{-1}
\end{align}
with ${\bf E}_{q} = \left[\left({\bf \bar{H}}_{q,q1} {\bf b}_{q1}\right) \dots \left({\bf \bar{H}}_{q,qN} {\bf b}_{qN}\right) \right] \in \mathbb{C}^{M_{\rm c} \times N}$ being the concatenated channels, after MRC, of the RRHs served by CU $q$. As in~\eqref{eq:beamformerAccess},  $\bm{\bar{\mu}}_q \in \mathbb{C}^{N \times N}$ is a diagonal matrix to meet the power constraint, having $\bar{\mu}_{qn}$ as its $n^{th}$ entry with (${\bf \widetilde{w}}_{qn}^{\rm{(b)}}$ is the $n^{th}$ column of $\widetilde{W}_q^{\rm{(b)}}$).
\begin{align}\label{eq:mu_qn}
\bar{\mu}_{qn} = \left[\bm{\bar{\mu}}_q\right]_{nn} = \sqrt{N^{-1} \mathbb{E}\left\{\|{\bf \widetilde{w}}_{qn}^{\rm{(b)}}\|^2 \right\}^{-1}}
\end{align}

\setlength{\textfloatsep}{30pt}

Finally, the spectral efficiency on the fronthaul is defined as
\vspace{-0.2em}
\begin{align}
&R_{qn}^{\rm{(b)}}(x_{qn},y_{qn})= \log\left(1+\frac{S_{qn}} {I_{qn} + \sigma^2} \right)
=
\resizebox{0.51\textwidth}{!}
{$\displaystyle
\log\left(1+\frac{\displaystyle p_{\rm c} \left| {\bf b}_{qn}^H {\bf \bar{H}}_{q,qn}^H {\bf \bar{w}}_{qn}^{\rm{(b)}} \right|^2} {
p_{\rm c}\sum_{q' \neq q} \sum_{n'= 1}^{N}\left| {\bf b}_{qn}^H {\bf \bar{H}}_{q',qn}^H {\bf \bar{w}}_{q'n'}^{\rm{(b)}}\right|^2 + \sigma_z^2} \right)
$}
\end{align}
Note that, for the chosen normalization, the desired signal power $S_{qn} = p_{\rm c} \bar{\mu}_{qn}^2$, is a constant determined by the RRH locations. Our goal is to develop an optimization problem similar to~\eqref{eq:multicast_singleRxOpt}, but for ZF; as with the multicast case, this requires the signal power (a constant) and a statistical characterization of the interference (using a gamma RV).  

\noindent \emph{Signal Power}: For the ZF scheme using the beamformer defined in~\eqref{eq:beamformerZFfronthaul} and MRC with $M$ receive antennas per RRH, the desired signal power is given by
{\allowdisplaybreaks
\begin{align}
S^{(b)}_{qn} &= p_{\rm c} \bar{\mu}_{qn}^2 = \frac{p_{\rm c}}{N} \mathbb{E}\left\{\|{\bf \widetilde{w}}_{qn}^{\rm{(b)}}\|^2 \right\}^{-1}
= \frac{p_{\rm c}}{N} \mathbb{E}\left\{ \left[ \left({\bf \widetilde{W}}_{q}^{\rm{(b)}}\right)^H {\bf \widetilde{W}}_{q}^{\rm{(b)}} \right]_{nn} \right\}^{-1}
\nonumber \\
&
= \frac{p_{\rm c}}{N} \mathbb{E}\left\{ \left[ \left( {\bf E}_{q}^H {\bf E}_{q} \right)^{-1} \right]_{nn} \right\}^{-1}
= \frac{p_{\rm c}}{N} \bar{\ell}_{q}(x_{qn},y_{qn}) \mathbb{E}\left\{ \left[ \left( \left({\bf E}_{q}^{(G)}\right)^H {\bf E}_{q}^{(G)} \right)^{-1} \right]_{nn} \right\}^{-1}
\nonumber \\
&
\stackrel{(a)}{=}
\frac{p_{\rm c}}{N} \bar{\ell}_{q}(x_{qn},y_{qn})  \left(\mathbb{E}\left\{\sigma_{q,qn}^{(G)} \right\}\right)^2 \mathbb{E}\left\{ \left[ \left( {\bm \Lambda}_q^H {\bm \Lambda}_q \right)^{-1} \right]_{nn} \right\}^{-1}
\nonumber \\
&
= 
\frac{p_{\rm c}}{N} \bar{\ell}_{q}(x_{qn},y_{qn}) \left(\mathbb{E}\left\{\sigma_{q,qn}^{(G)} \right\}\right)^2 N \mathbb{E}\left\{
\text{{\bf tr}}\left\{
\left( {\bm \Lambda}_q^H {\bm \Lambda}_q \right)^{-1} \right\} \right\}^{-1}
\end{align}
}
$\!\!$where ${\bf E}_{q}^{(G)} = \left[\left({\bf \bar{G}}_{q,q1} {\bf b}_{q1}\right) \dots \left({\bf \bar{G}}_{q,qN} {\bf b}_{qN}\right)\right] \in \mathbb{C}^{M_{\rm c} \times N}$ (${\bf \bar{G}}_{q,qn}$ represents the small-scale fading component of the corresponding channels, ${\bf \bar{H}}_{q,qn}$). Further, ${\bm \Lambda}_q = \left[ \left[{\bf U}_{q,q1}\right]_{.1} \dots \left[{\bf U}_{q,qN}\right]_{.1} \right] \in \mathbb{C}^{M_{\rm c} \times N}$ is the concatenation of the first columns of the unitary matrices ${\bf U}_{q,qn}, \forall n$ in cell $q$ resulting from the SVD of the channels ${\bf \bar{H}}_{q,qn}$. As for $(a)$, it follows from ${\bf \bar{G}}_{q,qn} {\bf b}_{qn} = \sigma_{q,qn}^{(G)} \left[{\bf U}_{q,qn}\right]_{.1}$.
\par 

The matrix ${\bm \Lambda}_q $ is random with statistically independent columns, but not Gaussian since $\left[{\bf U}_{q,qn}\right]_{.1}$ (the singular vector corresponding to the largest singular value of $\bar{\mathbf{H}}_{q,qn}$) is unit-norm. Not much is known about such matrices; for tractability, since $M_{\rm c}$ is large, we approximate $(M_{\rm c}-1) {\bm \Lambda}_q^H {\bm \Lambda}_q \sim \mathcal{W}_N\left(M_{\rm c},{\bf I}_{N}\right)$ as a $N \times N$ complex Wishart matrix with $M_{\rm c}$ degrees of freedom~\cite{1237134}. Therefore, $\mathbb{E}\left\{\text{{\bf tr}}\left\{\left({\bm \Lambda}_q^H{\bm \Lambda}_q\right)^{-1}\right\}\right\} \simeq (M_{\rm c}-1)\frac{N}{M_{\rm c} - N}$ (as we will see, this identity is extremely accurate). Consequently, the signal power is given by
\begin{align}
&S^{(b)}_{qn} \simeq
\frac{p_{\rm c}}{N} \bar{\ell}_{q}(x_{qn},y_{qn}) \left(\mathbb{E}\left\{\sigma_{q,qn}^{(G)} \right\}\right)^2 \frac{M_{\rm c} - N}{M_{\rm c} - 1}  \label{eq:ZFBFMeanSignal}
\end{align}
Note that, for $M_{\rm c} \gg N$, the term $\mathbb{E}\left\{ \left[ \left( {\bm \Lambda}_q^H {\bm \Lambda}_q \right)^{-1} \right]_{nn} \right\} \rightarrow 1$. As for the interference power, as before we need its mean and variance to define its gamma distribution. The mean is given by
{\allowdisplaybreaks
\begin{align}
\mathbb{E}\left\{I^{(b)}_{qn} \right\} &= p_{\rm c}\sum_{q' \neq q} \sum_{n'= 1}^{N} \mathbb{E}\left\{ | {\bf b}_{qn}^H {\bf \bar{H}}_{q',qn}^H {\bf \bar{w}}_{q'n'}^{\rm{(b)}}|^2 \right\}
= p_{\rm c}\sum_{q' \neq q} \sum_{n'= 1}^{N} \bar{\ell}_{q'}(x_{qn},y_{qn}) \mathbb{E}\left\{ | {\bf b}_{qn}^H {\bf \bar{G}}_{q',qn}^H {\bf \bar{w}}_{q'n'}^{\rm{(b)}}|^2 \right\}
\nonumber \\
&= p_{\rm c}\sum_{q' \neq q} \sum_{n'= 1}^{N} \bar{\ell}_{q'}(x_{qn},y_{qn})
\mathbb{E}\left\{ \left({\bf \bar{w}}_{q'n'}^{\rm{(b)}}\right)^H {\bf \bar{G}}_{q',qn} {\bf b}_{qn} {\bf b}_{qn}^H {\bf \bar{G}}_{q',qn}^H {\bf \bar{w}}_{q'n'}^{\rm{(b)}} \right\}
\nonumber \\
&\stackrel{(a)}{=} p_{\rm c}\sum_{q' \neq q} \sum_{n'= 1}^{N} \bar{\ell}_{q'}(x_{qn},y_{qn}) \bar{\mu}_{q'n'}^2
\mathbb{E}\left\{ \left({\bf \widetilde{w}}_{q'n'}^{\rm{(b)}}\right)^H {\bf \bar{G}}_{q',qn} {\bf b}_{qn} {\bf b}_{qn}^H {\bf \bar{G}}_{q',qn}^H {\bf \widetilde{w}}_{q'n'}^{\rm{(b)}} \right\}
\nonumber \\
&= p_{\rm c}\sum_{q' \neq q} \sum_{n'= 1}^{N} \bar{\ell}_{q'}(x_{qn},y_{qn}) \bar{\mu}_{q'n'}^2
\mathbb{E}\left\{ \text{\textbf{tr}}\left\{
{\bf \bar{G}}_{q',qn} {\bf b}_{qn} {\bf b}_{qn}^H {\bf \bar{G}}_{q',qn}^H {\bf \widetilde{w}}_{q'n'}^{\rm{(b)}} \left({\bf \widetilde{w}}_{q'n'}^{\rm{(b)}}\right)^H \right\}
\right\}
\nonumber \\
& \stackrel{(b)}{=} p_{\rm c}\sum_{q' \neq q} \sum_{n'= 1}^{N} \bar{\ell}_{q'}(x_{qn},y_{qn}) \bar{\mu}_{q'n'}^2
\mathbb{E}\left\{ \text{\textbf{tr}}\left\{
{\bf b}_{qn}^H {\bf \bar{G}}_{q',qn}^H \mathbb{E}\left\{ {\bf \widetilde{w}}_{q'n'}^{\rm{(b)}} \left({\bf \widetilde{w}}_{q'n'}^{\rm{(b)}}\right)^H \right\}
{\bf \bar{G}}_{q',qn} {\bf b}_{qn} \right\}
\right\}
\nonumber \\
&\stackrel{(c)}{=} p_{\rm c}\sum_{q' \neq q} \sum_{n'= 1}^{N} \bar{\ell}_{q'}(x_{qn},y_{qn}) \bar{\mu}_{q'n'}^2
\mathbb{E}\left\{ \| {\bf \widetilde{w}}_{q'n'}^{\rm{(b)}} \|^2 \right\}
\nonumber \\
&
= p_{\rm c}\sum_{q' \neq q} \sum_{n'= 1}^{N} \bar{\ell}_{q'}(x_{qn},y_{qn})
{N}^{-1}
= p_{\rm c}\sum_{q' \neq q} \bar{\ell}_{q'}(x_{qn},y_{qn}) \label{eq:ZFBFIqnMean}
\end{align}
}
$\!\!$where $(a)$ follows from the fact that $\bm{\bar{\mu}}_{q'}$ is a diagonal matrix, hence ${\bf \bar{w}}_{q'n'}^{\rm{(b)}} = \left[ {\bf \bar{W}}_{q'}^{\rm{(b)}} \right]_{.n'} = {\bf \widetilde{W}}_{q'}^{\rm{(b)}} \left[ \bm{\bar{\mu}}_{q'} \right]_{.n'} = \bar{\mu}_{q'n'} {\bf \widetilde{w}}_{q'n'}^{\rm{(b)}}$, $(b)$ follows from the cyclic property of the trace and from the independence of ${\bf \widetilde{w}}_{q'n'}^{\rm{(b)}}$, ${\bf \bar{G}}_{q',qn}$ and ${\bf b}_{qn}$, and $(c)$ follows from $\mathbb{E}\left\{ {\bf \bar{G}}_{q',qn} {\bf b}_{qn} {\bf b}_{qn}^H {\bf \bar{G}}_{q',qn}^H \right\} = {\bf I}_{M_{\rm c}}$. 

For the variance we have
\begin{align}
\mathtt{Var}\left\{I^{(b)}_{qn} \right\}
&= 
\resizebox{0.5\textwidth}{!}
{$\displaystyle
\sum_{q' \neq q} \sum_{n'= 1}^{N} \left(p_{\rm c} \bar{\ell}_{q'}(x_{qn},y_{qn})\right)^2 
\mathtt{Var}\left\{ | {\bf b}_{qn}^H {\bf \bar{G}}_{q',qn}^H {\bf \bar{w}}_{q'n'}^{\rm{(b)}}|^2 \right\}
$}
\stackrel{(a)}{=}
\resizebox{0.25\textwidth}{!}
{$\displaystyle
N^{-1} \sum_{q' \neq q} \left(p_{\rm c} \bar{\ell}_{q'}(x_{qn},y_{qn})\right)^2
$}
\label{eq:ZFBFIqnVariance}
\end{align}
where $(a)$ follows from the fact that $X = \left| {\bf b}_{qn}^H {\bf \bar{G}}_{q',qn}^H {\bf \bar{w}}_{q'n'}^{\rm{(b)}}\right|^2$ is the norm of a complex Gaussian RV, and hence, has an exponential distribution with mean $1/N$, i.e., its variance is $1/N^2$. The $N$ terms in the summation (over $n'$) leads to the $N^{-1}$ in~\eqref{eq:ZFBFIqnVariance}.

\noindent \textit{RRH Location Optimization:} We are now ready to state the resulting optimization problem for RRH locations. Our optimization problem is to, again, maximize the typical access rate given a fronthaul outage constraint. As in Section~\ref{subsec:PlacementMulticastAlg}, we use block coordinate descent and focus on a specific cell $q$ keeping the RRH positions in the other cells fixed (and then iterate over cells). For cell $q$, the optimization problem is
\begin{subequations}\label{eq:opt3Formulation}
	\begin{align}
	&\underset{{\bf x}_q,{\bf y}_q}{\max}\quad \mathbb{E}_{\widetilde{x}_k,\widetilde{y}_k}\left\{ R_{qk}^{\rm{(a)}}({\bf x},{\bf y}) \right\},
	\label{eq:ObjectiveFunct1}\\
	&\text{s.t.}\quad
	%
	%
	\displaystyle \mathbb{P}\left\{R_{qn}^{\rm{(b)}}(x_{qn},y_{qn}) \le K \frac{\omega}{\omega_{\rm b} }
	\mathbb{E}_{\widetilde{x}_k,\widetilde{y}_k}\left\{ R_{qk}^{\rm{(a)}}({\bf x},{\bf y}) \right\}
	\right\}\le\epsilon,
	%
	\quad n=1,\dots,N.
	\label{eq:Opt3_Constraint1}
	\end{align}
\end{subequations}

As before, we model the interference power as a gamma RV with mean and variance given in~\eqref{eq:ZFBFIqnMean} and~\eqref{eq:ZFBFIqnVariance} respectively; the signal power is a constant given in~\eqref{eq:ZFBFMeanSignal}; we can rewrite the constraint in \eqref{eq:Opt3_Constraint1} as
\begin{align}\label{eq:ZF-BF_outage}
1 - F_{I^{(b)}_{qn}}\bigg(
\frac{ S^{(b)}_{qn} }{e^{\dot{\mathsf{x}}} - 1} - \sigma_z^2
\bigg)
\le\epsilon, \quad n=1,\dots,N
\end{align}
where $\dot{\mathsf{x}}$ is defined below~\eqref{eq:multicastfronthaulOutage}. The RRH locations are again optimized using Algorithm~\ref{algorithm:iterativeAlgoMulticast}, with the appropriate change in the choice of constraint, namely, in Step~\ref{step:iterativeAlgoMulticast_constraintCheck} in Algorithm~\ref{algorithm:iterativeAlgoMulticast}, we check the validity of the constraint in~\eqref{eq:ZF-BF_outage} instead of the constraint in~\eqref{eq:multicast_singleRxOpt}.

\section{Simulations and Results}\label{sec:results}
In this section, we present numerical results illustrating the efficiency of our analysis, our approximations, and Algorithm~\ref{algorithm:iterativeAlgoMulticast}. We consider a network of $Q=9$ square cells of area $1~{\rm km}^2$ each, with wraparound (our approach is general and can be applied to any cellular network). We consider a system of $25$ RBs used for the data plane, each with a bandwidth of $180~{\rm kHz}$. 

To model the traffic distribution, similar to~\cite{7277444}, we use a flexible PDF combining a uniform distribution chosen with probability $P_0$, and a random number, $N_h$, of bivariate normal distributions representing hotspots chosen with probability $P_0$. The value of $P_0$ can be used to control the density of the hotspots. The number $N_h$ of these hotspots is uniformly distributed in the interval $\left[N_h^\text{min}, N_h^\text{max} \right]$. These hotspots are centered at locations $\left[{\bf \widetilde{x}}_h,{\bf \widetilde{y}}_h\right] = \left[\left[{\widetilde{x}}_{h1},\dots,{\widetilde{x}}_{h N_h}\right]^T, \left[{\widetilde{y}}_{h1}, \dots, {\widetilde{y}}_{h N_h}\right]^T\right] \in \mathbb{R}^{N_h\times 2}$, and they have equal variances $\sigma_h^2$ in both dimensions, without any correlation between the two axes. Therefore, the traffic distribution PDF in a cell bounded by the cell boundary $\mathcal{B}_{q}$ is defined as
	\begin{align}\label{eq:trafficDist}
		\resizebox{0.92\textwidth}{!}
		{$\displaystyle
			f_q(\widetilde{x}_k,\widetilde{y}_k) = C \left(P_0 \left( \frac{1}{B_{q}} \right)
			+ \left( 1 - P_0 \right)
			\frac{1}{N_h 2 \pi \sigma_h^2} \sum_{i = 1}^{N_h} \left( \exp\left(-\frac{\left(\widetilde{x}_k - \widetilde{x}_{h_i}\right)^2+\left(\widetilde{y}_k - \widetilde{y}_{h_i}\right)^2}{2\sigma_h^2}\right) \right) \right),
			$}
	\end{align}
where $B_{q}$ (different from cell boundary $\mathcal{B}_{q}$) is the area of cell $q$ and $C$ is a factor to normalize the PDF. We note that any PDF derived from, for example, a traffic survey can be used.

\begin{table}[t]
	\scriptsize
	\centering
	\begin{tabular}{|p{0.08\linewidth}|p{0.13\linewidth}|p{0.18\linewidth}||p{0.08\linewidth}|p{0.14\linewidth}|p{0.22\linewidth}|}
		\hline
		\hline
		\multicolumn{1}{|l|}{ \textit{\textbf{Description}}} & \multicolumn{1}{l|}{ \textit{\textbf{Parameter}}} & \multicolumn{1}{l||}{\textit{\textbf{Value}}}& \multicolumn{1}{l|}{ \textit{\textbf{Description}}} & \multicolumn{1}{l|}{ \textit{\textbf{Parameter}}} & \multicolumn{1}{l|}{\textit{\textbf{Value}}}\\
		\hline
		Cell config. & $Q$, $N$, $M_{\rm c}$, $M$, $K$ & $9$, $10$, $64$, $8$, $10$ & Hotspots & \pbox{20cm}{$P_0$, $\sigma_h$; $N_h^\text{min}$, $N_h^\text{max}$} & \pbox{20cm}{$0.1$,$100~{\rm m}$; Per network:$2Q$,$4Q$} \\
		\hline
		Power & $p$, $p_{\rm c}$  & $30~{\rm dBm}$, $45~{\rm dBm}$ & Algorithm & $\epsilon$, $d_\text{cvg}$, $\nu_f$, $\nu$ & $0.2$, $1~{\rm m}$, $0.9$, $10^6$ \\
		\hline
		Bandwidth &RB, $\omega$, $\omega_{\rm b}$ & $180~{\rm KHz}$, $5$ RBs, $20$ RBs & Path loss & $d_0$, $\alpha$ & $0.392~{\rm m}$, $3.76$ \\
		\hline
		Noise & $S_z$, $F_z$ & $-174~{\rm dBm/ Hz}$, $8~{\rm dBm}$ & \multicolumn{3}{c|}{}\\
		\hline
		\hline
	\end{tabular}
	\vspace{-0.5em}
	\caption{Simulation parameters.}
	\label{table:sim_parameters}   
	\vspace{-3em}
\end{table}

In these simulations, we ignore shadowing, hence considering homogeneous propagation environments. For non-homogeneous environments, different path loss models can be used in marked regions (can be a system input) in the network to account for the shadowing based on the location $(\widetilde{x}_k, \widetilde{y}_k)$ and the environment properties. Given the limitations on available space, we do not study this scenario.

Table~\ref{table:sim_parameters} lists the other parameters used in our simulations,  unless otherwise specified, including those for the hotspots. We assume that the available bandwidth is divided, without overhead, between the fronthaul and the access channel maintaining $\omega + \omega_{\rm b} = 25$ RBs.

\begin{figure}[t]
	\centering
	\begin{subfigure}[t]{0.32\textwidth}
		\raggedright
		\includegraphics[width=0.95\textwidth]{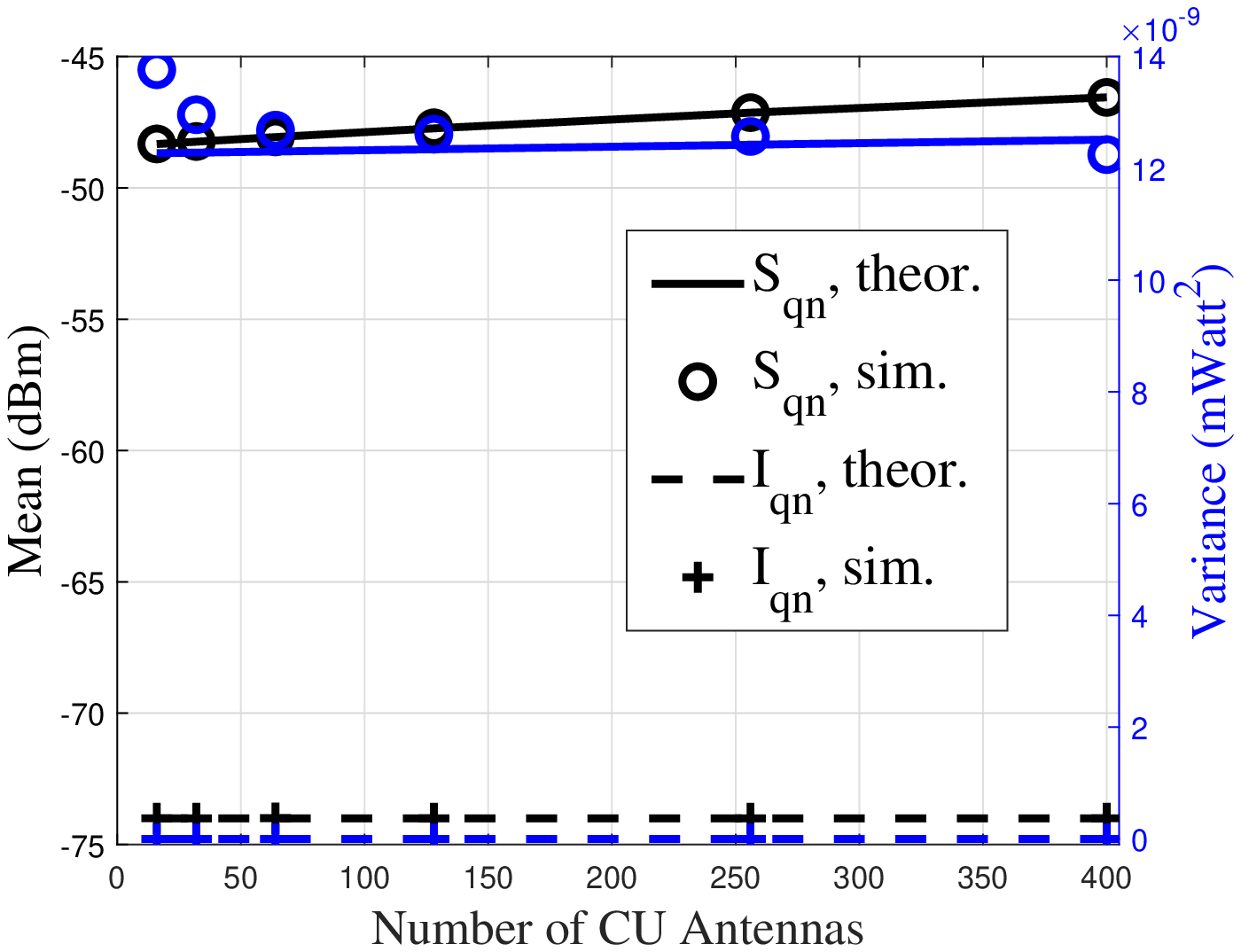}
		\caption{\raggedright Mean/variance obtained from \eqref{eq:SigfronthaulMean_Multicast}-\eqref{eq:IntfronthaulVar_Multicast} and simulations.}
		\label{fig:multicastMeanVariance}
	\end{subfigure}
	%
	%
	\begin{subfigure}[t]{0.32\textwidth}
		\centering
		\includegraphics[width=0.95\textwidth]{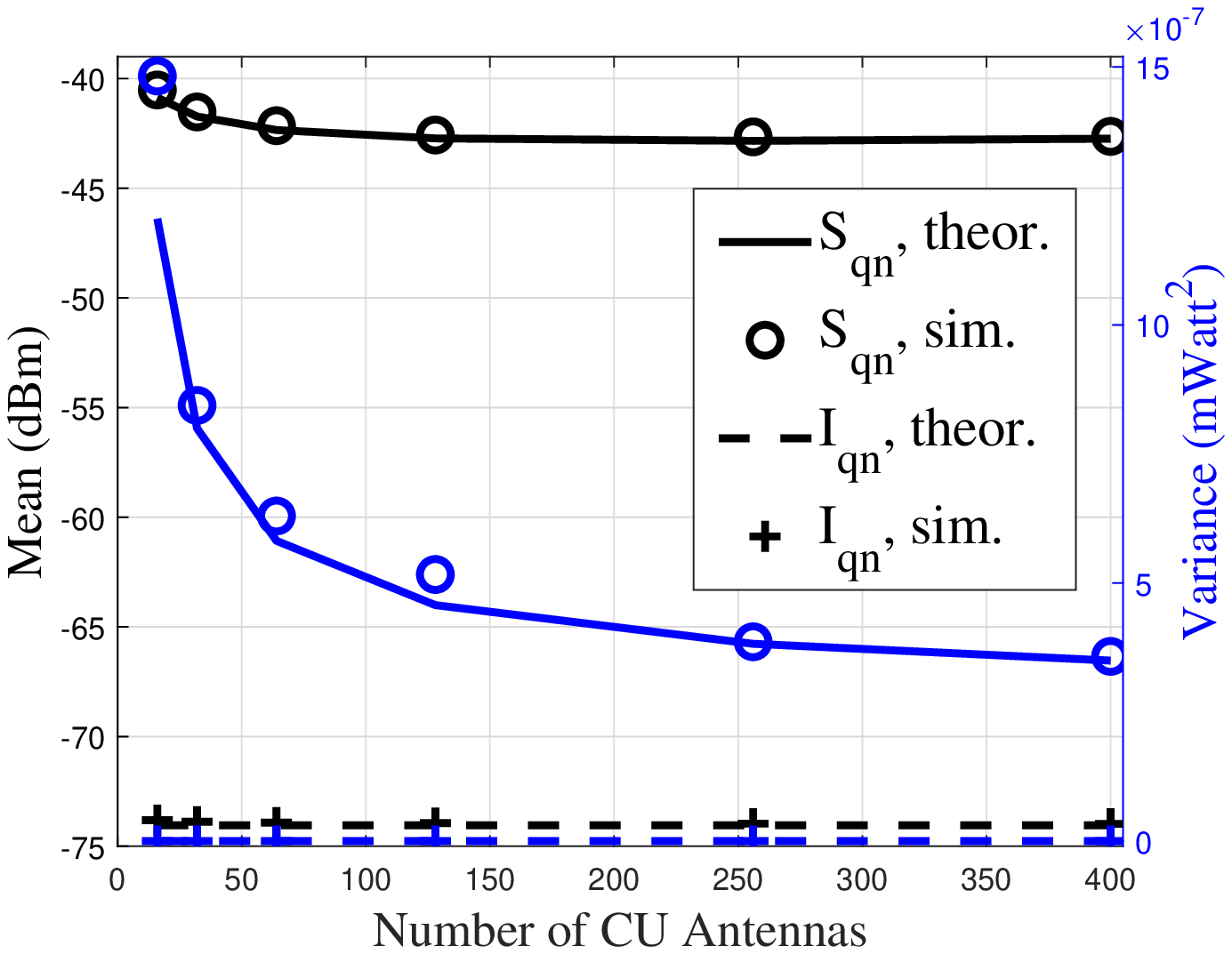}
		\caption{Mean/variance obtained from \eqref{eq:MultAntSigMean}-\eqref{eq:MultAntIntVar} and simulations.}
		\label{fig:multicastMeanVariance_MRC}
	\end{subfigure}
	%
	%
	\begin{subfigure}[t]{0.32\textwidth}
		\raggedleft
		\includegraphics[width=0.95\textwidth]{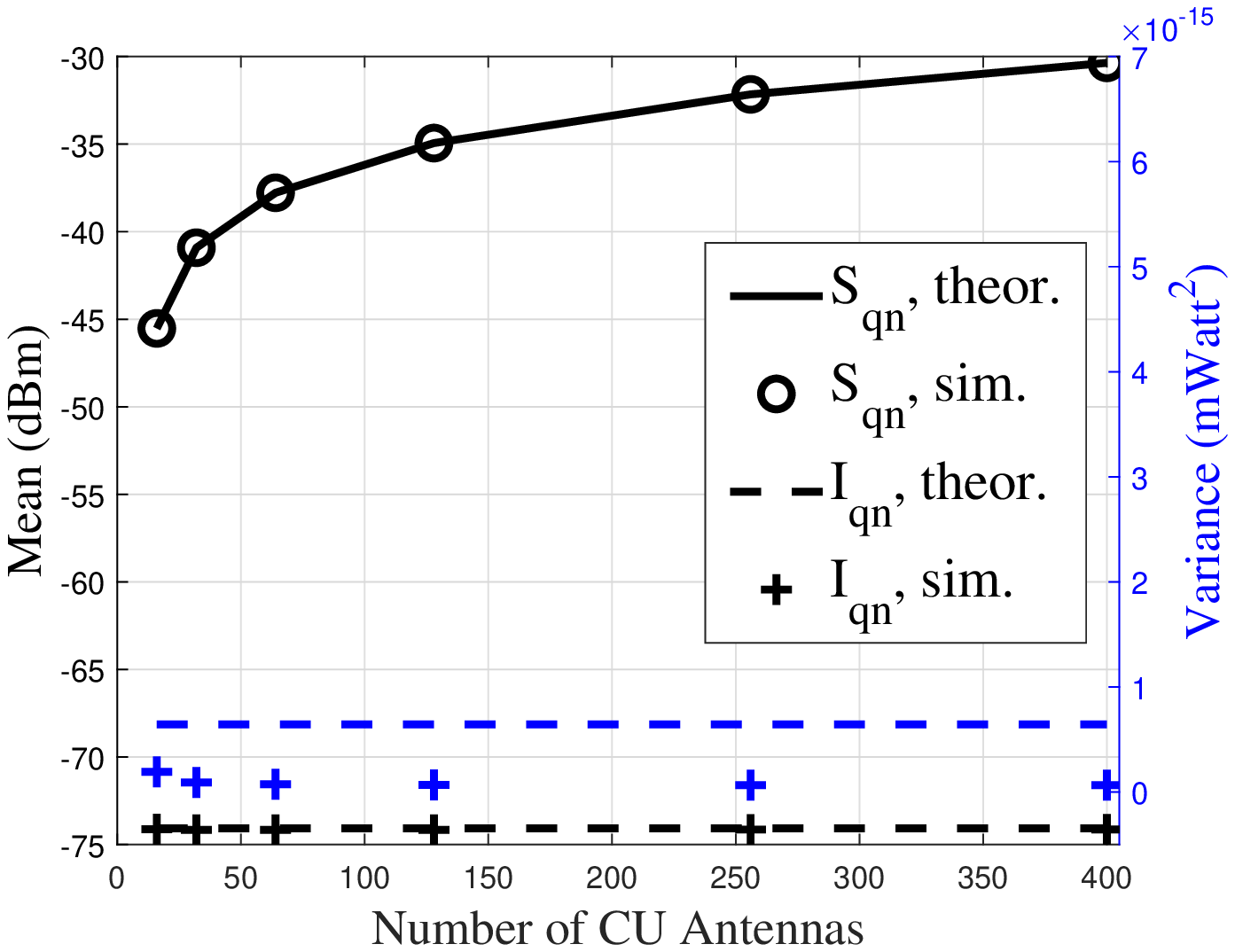}
		\caption{\raggedleft Mean/variance obtained from \eqref{eq:ZFBFMeanSignal}-\eqref{eq:ZFBFIqnVariance} and simulations.}
		\label{fig:ZFBFMeanVariance_MRC}
	\end{subfigure}
	\\
	\begin{subfigure}[t]{0.32\textwidth}
		\raggedright
		\includegraphics[width=0.89\textwidth]{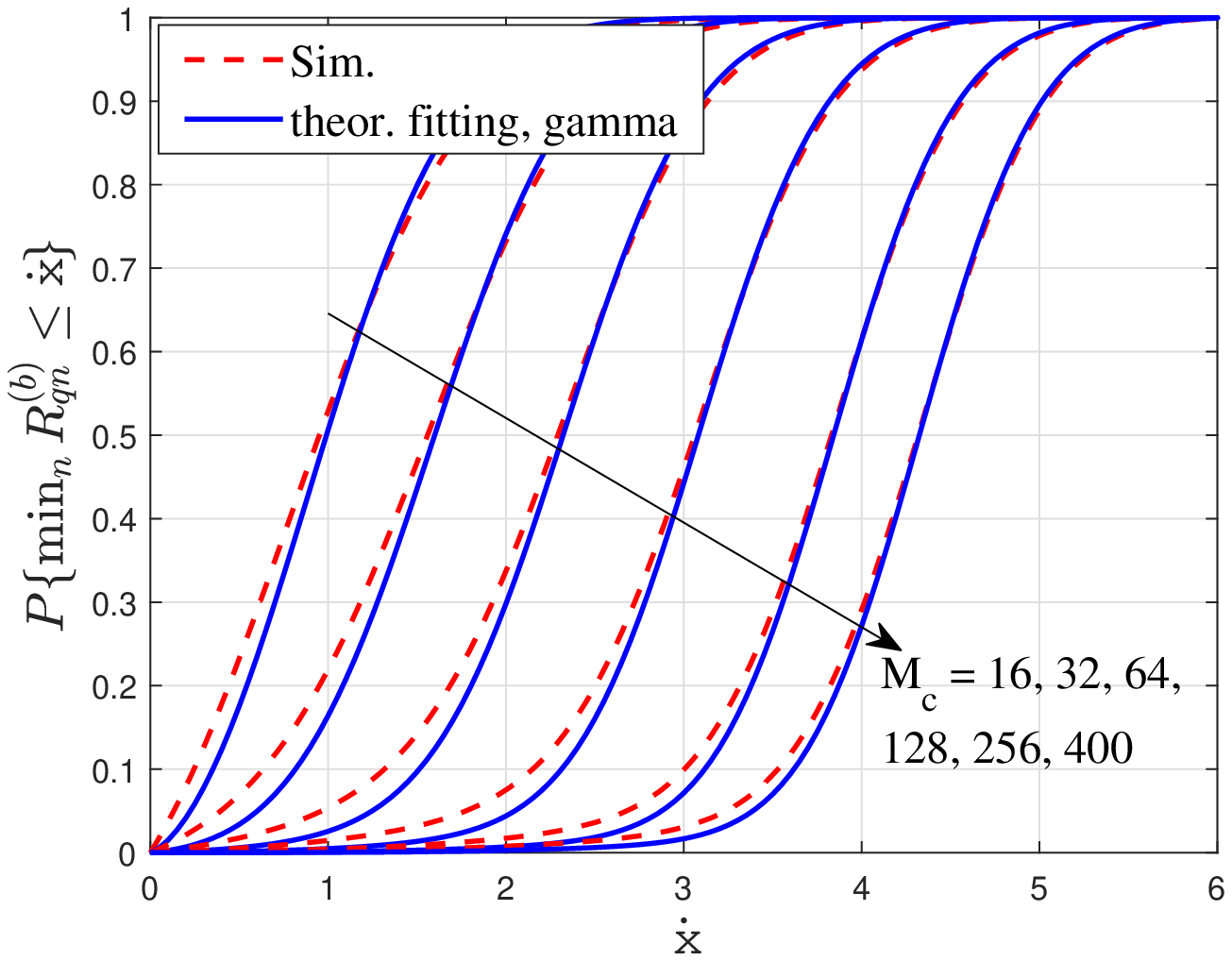}
		\caption{\raggedright Multicast fronthaul constraint obtained theoretically and from simulation. Single receive antenna.}
		\label{fig:multicastConstraint}
	\end{subfigure}
	\begin{subfigure}[t]{0.32\textwidth}
		\centering
		\includegraphics[width=0.89\textwidth]{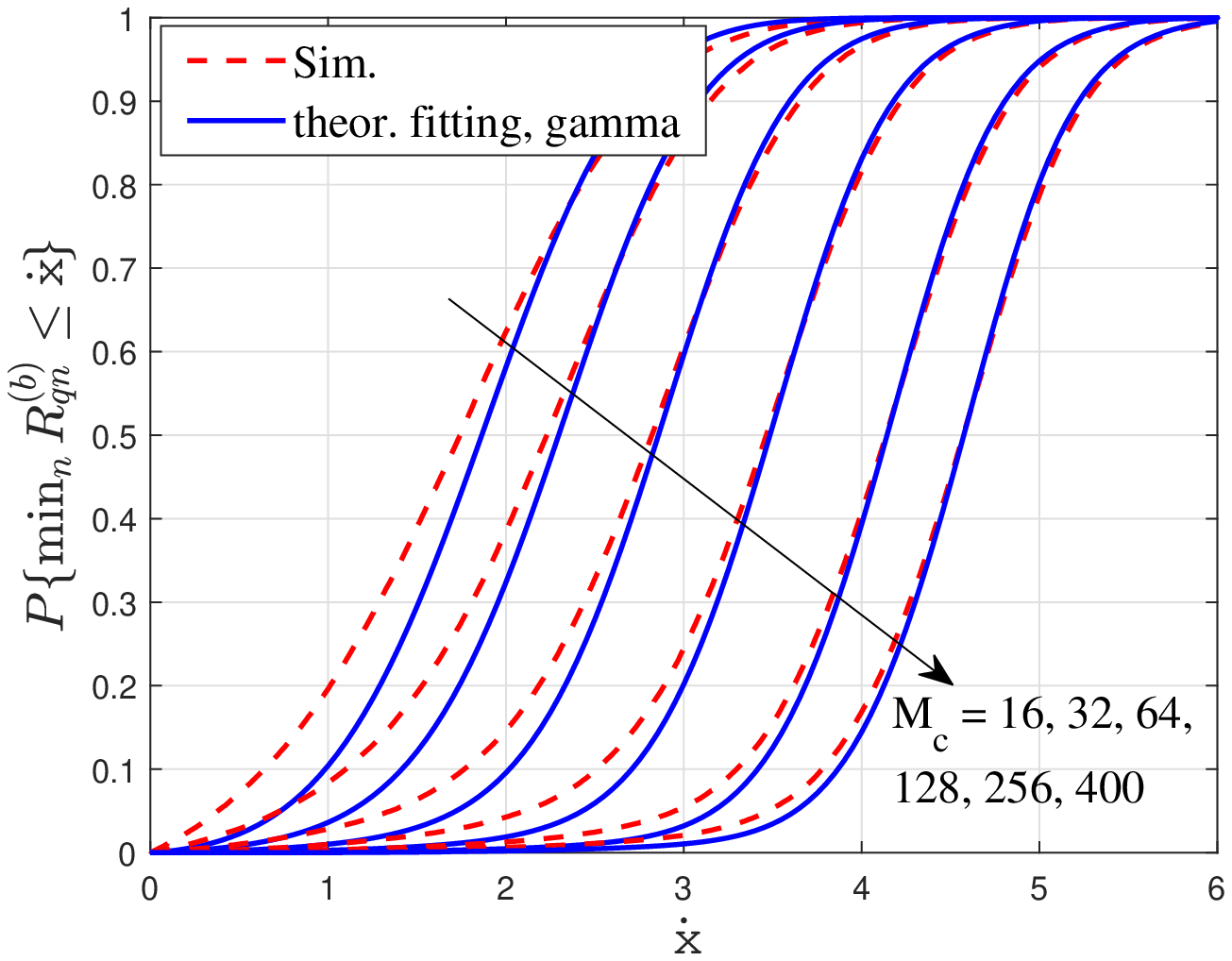}
		\caption{Multicast fronthaul constraint obtained theoretically and from simulation. Multiple receive antennas.}
		\label{fig:multicastConstraint_MRC}
	\end{subfigure}
	\begin{subfigure}[t]{0.32\textwidth}
		\raggedleft
		\includegraphics[width=0.89\textwidth]{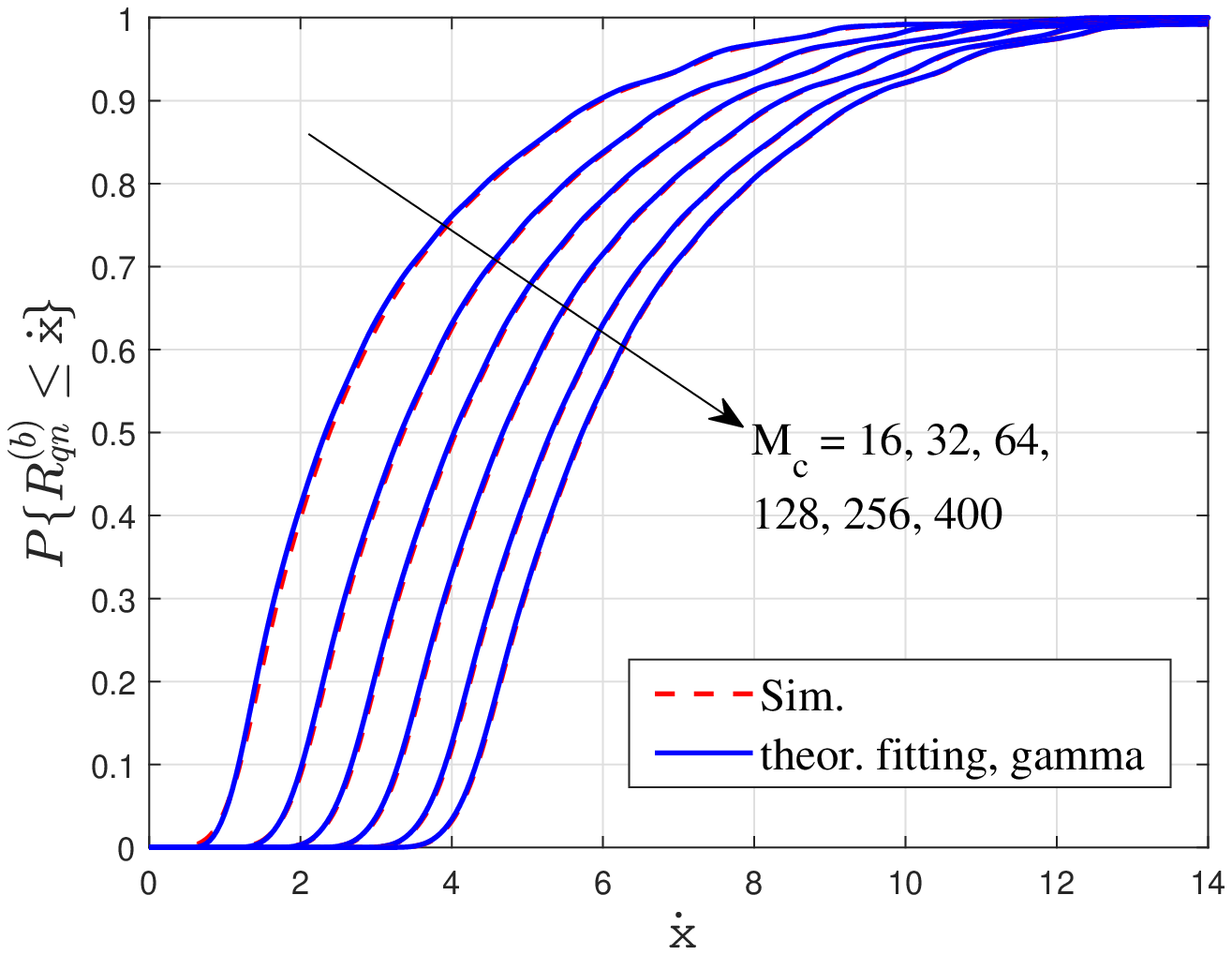}
		\caption{ZFBF fronthaul constraint obtained theoretically and from simulation. Multiple receive antennas.}
		\label{fig:ZFBFConstraint_MRC}
	\end{subfigure}
	\vspace{-0.5em}
	\caption{Illustrating the accuracy of the analysis of fronthaul MIMO schemes.}
	\label{fig:fronthaul_accuracy}
	\vspace{-1em}
\end{figure}

\noindent \textit{Gamma RV Approximation}: We begin by evaluating the accuracy of our constraint approximation in~\eqref{eq:Opt2Constraint1} for the multicast fronthaul scheme using~\eqref{eq:SigfronthaulMean_Multicast}-\eqref{eq:IntfronthaulVar_Multicast} (single receive antenna),~\eqref{eq:MultAntSigMean}-\eqref{eq:MultAntIntVar} (multiple receive antenna) and that in~\eqref{eq:ZF-BF_outage} for the ZF beamforming fronthaul using~\eqref{eq:ZFBFMeanSignal}-\eqref{eq:ZFBFIqnVariance}. We present the results of Monte Carlo simulations based on $100$ realizations of RRH locations, and a $1000$ small-scale fading channels for each location realization. The results in Figs.~\ref{fig:fronthaul_accuracy}(\subref{fig:multicastMeanVariance}),~\ref{fig:fronthaul_accuracy}(\subref{fig:multicastMeanVariance_MRC}) and~\ref{fig:fronthaul_accuracy}(\subref{fig:ZFBFMeanVariance_MRC}) compare the empirical mean and variance as a function of the number of antennas at the CU. The figures show that each expression is quite accurate for our purpose of system design. 

\begin{figure}[t]
	\centering
	\begin{minipage}{.46\textwidth}
		\includegraphics[width=0.95\textwidth]{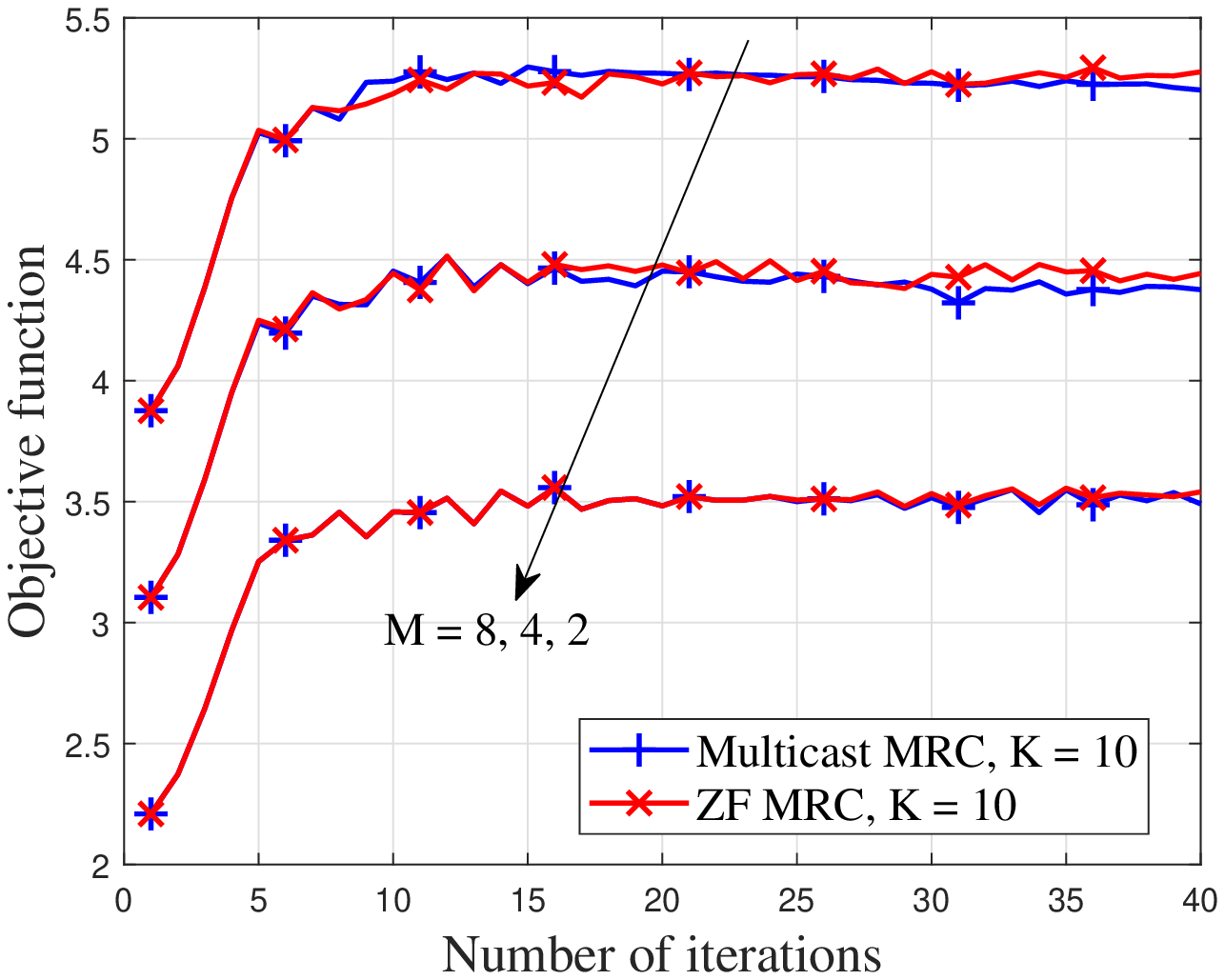}
		\vspace{-0.5em}
		\caption{Objective function vs number of iterations for the algorithm under different configurations.}  
		\label{fig:algo_iterN10_avgOver_RB1_till_6}
	\end{minipage}%
	\quad \quad
	\begin{minipage}{.46\textwidth}
		\centering
		\includegraphics[width=0.95\textwidth]{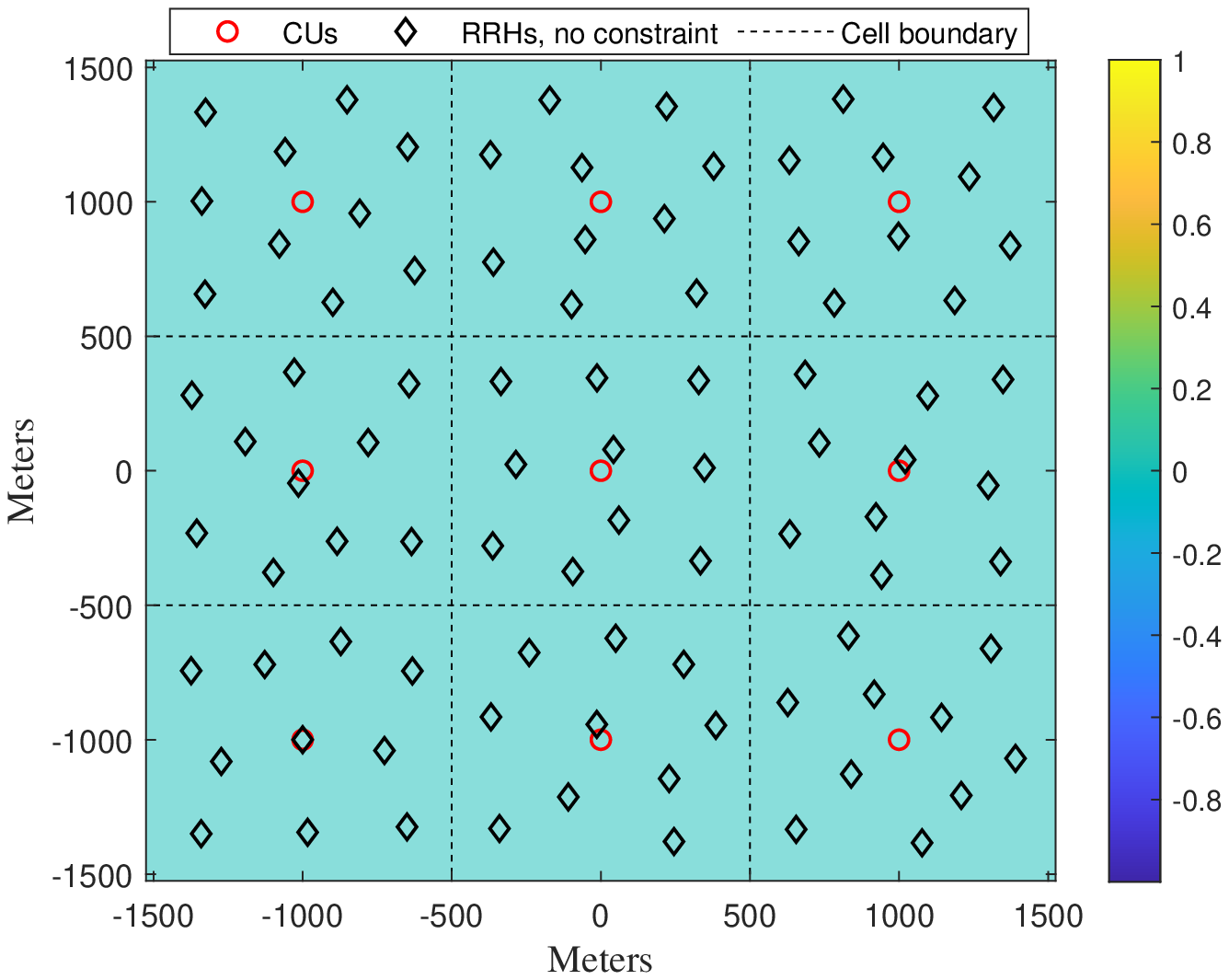}
		\vspace{-0.5em}
		\caption{RRHs deployed locations with no hotspots ($P_0 = 1$) and with a relaxed fronthaul constraint.}
		\label{fig:NoHotSpots_Test_Relaxed}
	\end{minipage}%
	\vspace{-1em}
\end{figure}

In Figs.~\ref{fig:fronthaul_accuracy}(\subref{fig:multicastConstraint}),~\ref{fig:fronthaul_accuracy}(\subref{fig:multicastConstraint_MRC}), and~\ref{fig:fronthaul_accuracy}(\subref{fig:ZFBFConstraint_MRC}) we compare the empirical and theoretical CDFs, effectively the empirical fronthaul outage versus a target rate ($x$-axis) and the one resulting from our gamma distribution fitting approach. Again, the empirical and theoretical outage rates match quite well; the small discrepancy between the two curves is explained by our usage of the gamma distribution, not the true, but unknown, distribution. The only exception is Fig.~\ref{fig:fronthaul_accuracy}(\subref{fig:multicastConstraint_MRC}), which suggests that by using the gamma distribution at small outage values we obtain an \emph{upperbound for the performance because we are underestimating the outage}. For the ZF fronthaul, we are characterizing the average performance (Fig.~\ref{fig:ZFBFConstraint_MRC}). The implication is that our observations discussed next are strengthened. The keypoint is that, because we use closed-form expressions and not numerical methods, we can now write an explicit expression for the fronthaul outage, which can, in turn, be used in our optimization problems.

\noindent \textit{Convergence:} In Fig.~\ref{fig:algo_iterN10_avgOver_RB1_till_6}, we plot the evolution of the objective function versus the number of iterations of the algorithm. One iteration represents a single update for the locations of RRHs in all cells. We present the convergence for the two different schemes, multicast with MRC and ZF with MRC, for different network configurations, where the plots show a typical convergence curve for one realization. The results show that while the objective function is not guaranteed to increase with every iteration, the algorithm converges in about $10$-$15$ iterations on average.

\subsection{RRH Location Optimization with Multicast Fronthaul}
In Fig.~\ref{fig:NoHotSpots_Test_Relaxed}, we show the obtained deployed locations for the RRHs when we have no hotspots in the network and the traffic distribution is uniformly distributed. This scenario can be realized by setting $P_0 = 1$ in the PDF of the traffic distribution in~\eqref{eq:trafficDist}. As expected, a relaxed fronthaul constraint results in a uniformly distributed deployment of the RRHs in the network area.

\begin{figure}[t]
	\begin{minipage}{.46\textwidth}
		\centering
		\includegraphics[width=0.95\textwidth]{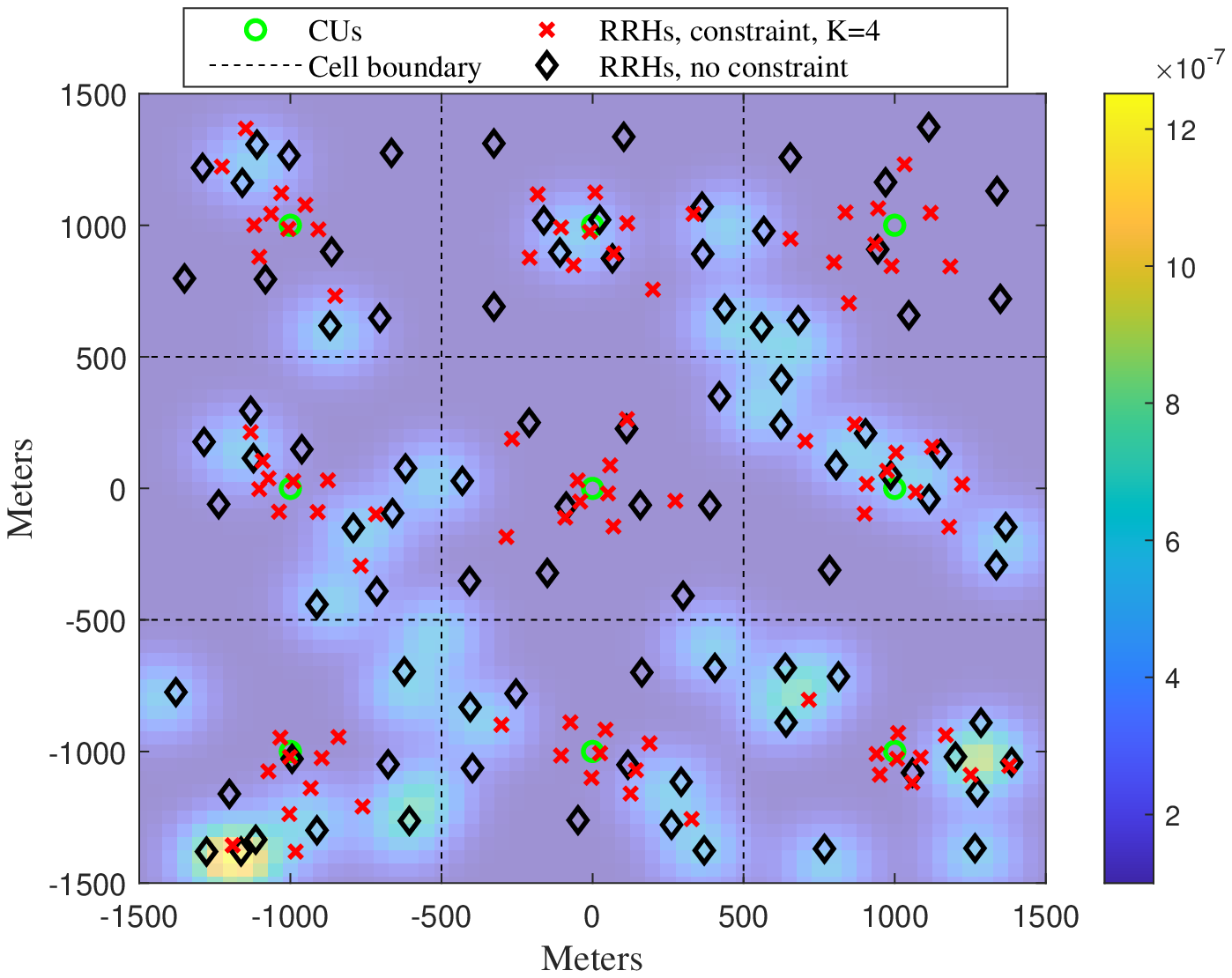}
		\vspace{-0.5em}
		\captionof{figure}{RRHs deployed locations using multicast fronthaul. $M = 8,\ \omega = 4~\text{RBs}$. $K =4$.}
		\label{fig:DeploymentMulticastK4}
	\end{minipage}%
	\quad\quad
	\begin{minipage}{.46\textwidth}
		\centering
		\includegraphics[width=0.95\textwidth]{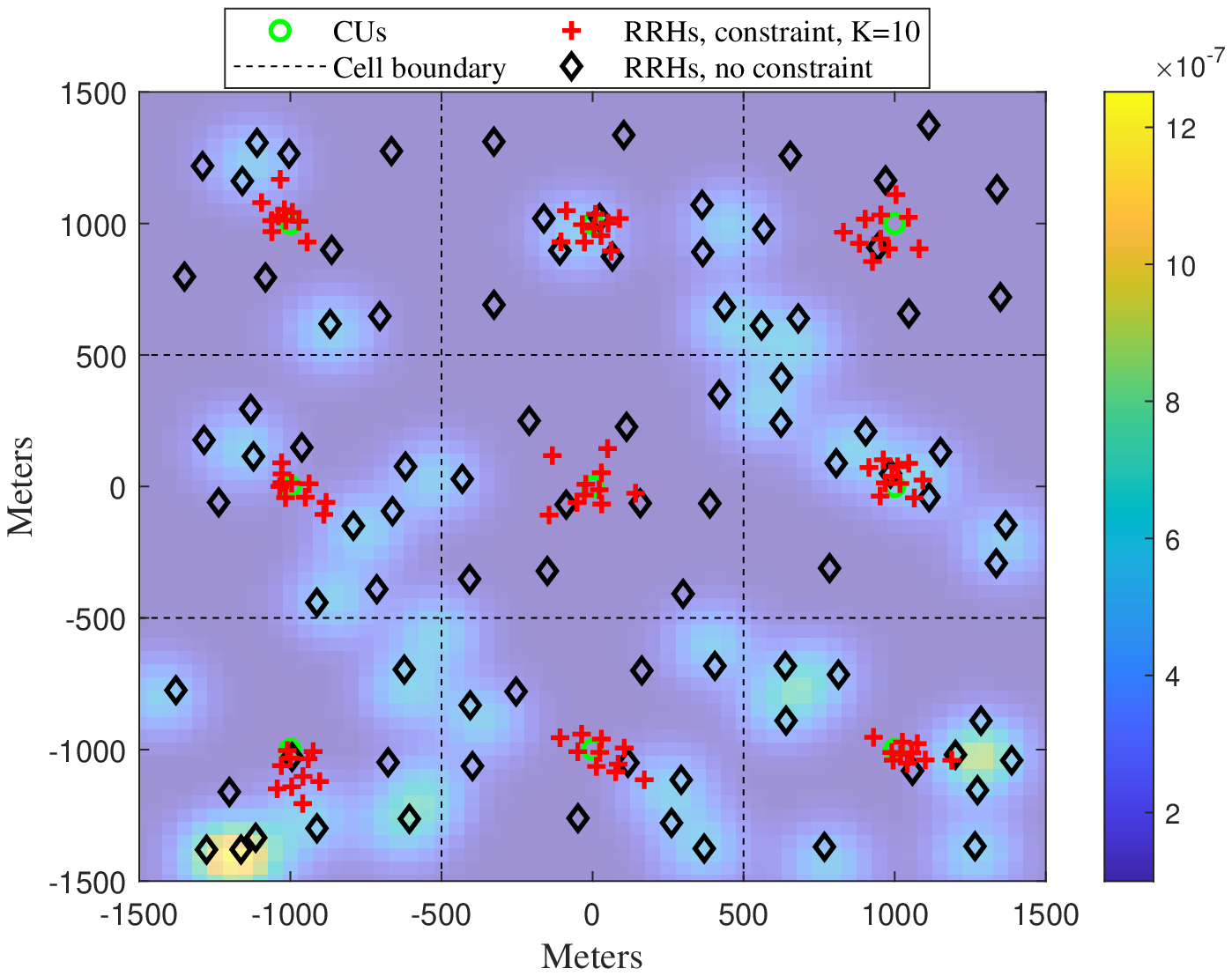}		
		\vspace{-0.5em}
		\captionof{figure}{RRHs deployed locations using multicast fronthaul $M = 8,\ \omega = 4~\text{RBs}$. $K = 10$}
		\label{fig:DeploymentMulticastK10}
	\end{minipage}%
	\vspace{-1.5em}
\end{figure}

In each of Figs.~\ref{fig:DeploymentMulticastK4} and~\ref{fig:DeploymentMulticastK10} we plot different resulting RRH deployments for the scenario upon using Algorithm~\ref{algorithm:iterativeAlgoMulticast} with $M = 8$ antennas per RRH and $\omega = 4$ RBs for the access channel. Here, the colormap indicates the traffic density. The figures show deployments without a fronthaul constraint and with the fronthaul constraint using either $K=4$ or $K=10$. They also illustrate two key issues: first, comparing the case with and without the fronthaul constraint, we see that the fronthaul plays a significant role in where RRHs can be deployed. Second, serving a large number of users simultaneously ($K=10$) significantly constrains the locations of the RRHs. The implication is that attempting to serve a large number of users restricts the RRHs to being close to the CU, i.e., bringing into question the distributed nature of the network.

\begin{figure}[t]
	\begin{minipage}{.32\textwidth}
		\raggedright
		\includegraphics[width=1\textwidth]{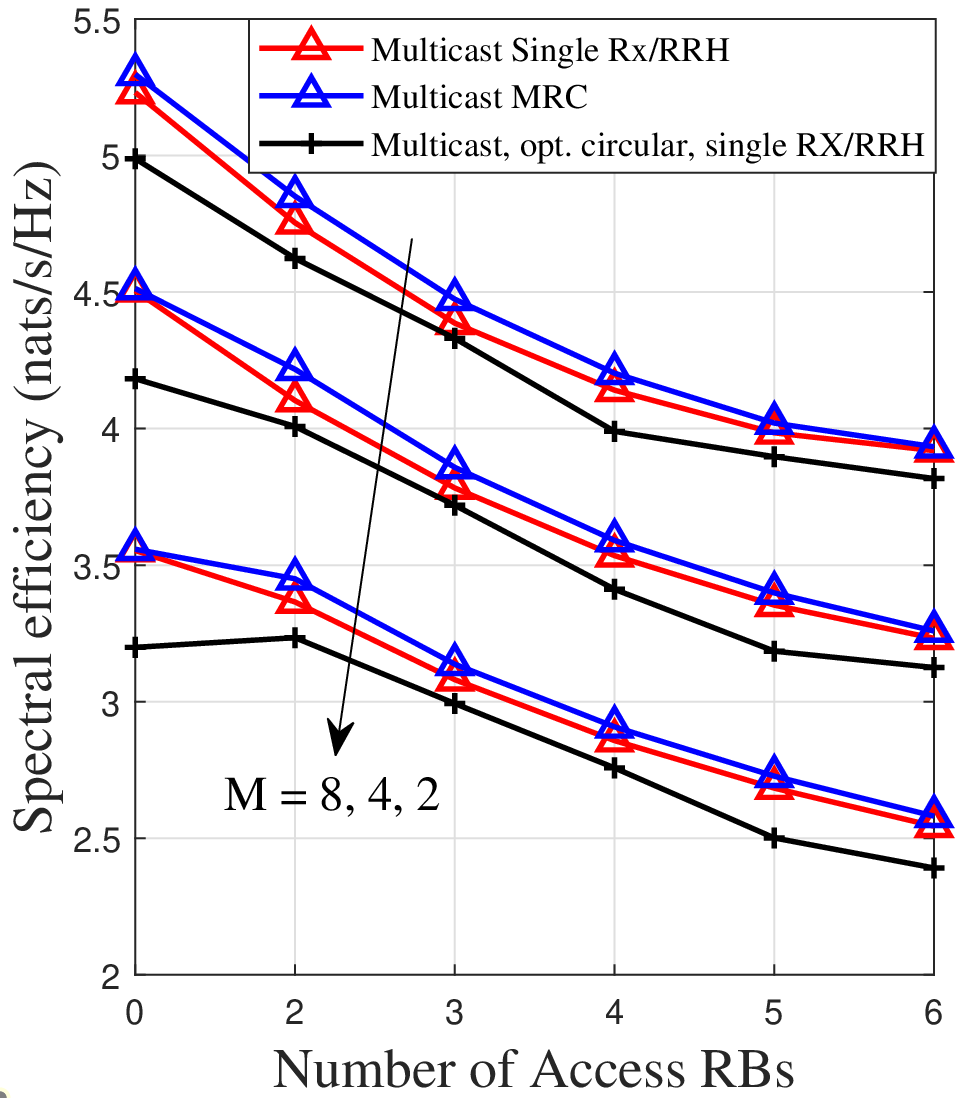}
		\vspace{-2em}
		\caption{Achieved spectral efficiency for different RB access configuration at $K=10$.}
		\label{fig:spectralEfficiency_multicast_sameNet}
	\end{minipage}%
	$\ $
	\begin{minipage}{.32\textwidth}
		\centering
		\includegraphics[width=1\textwidth]{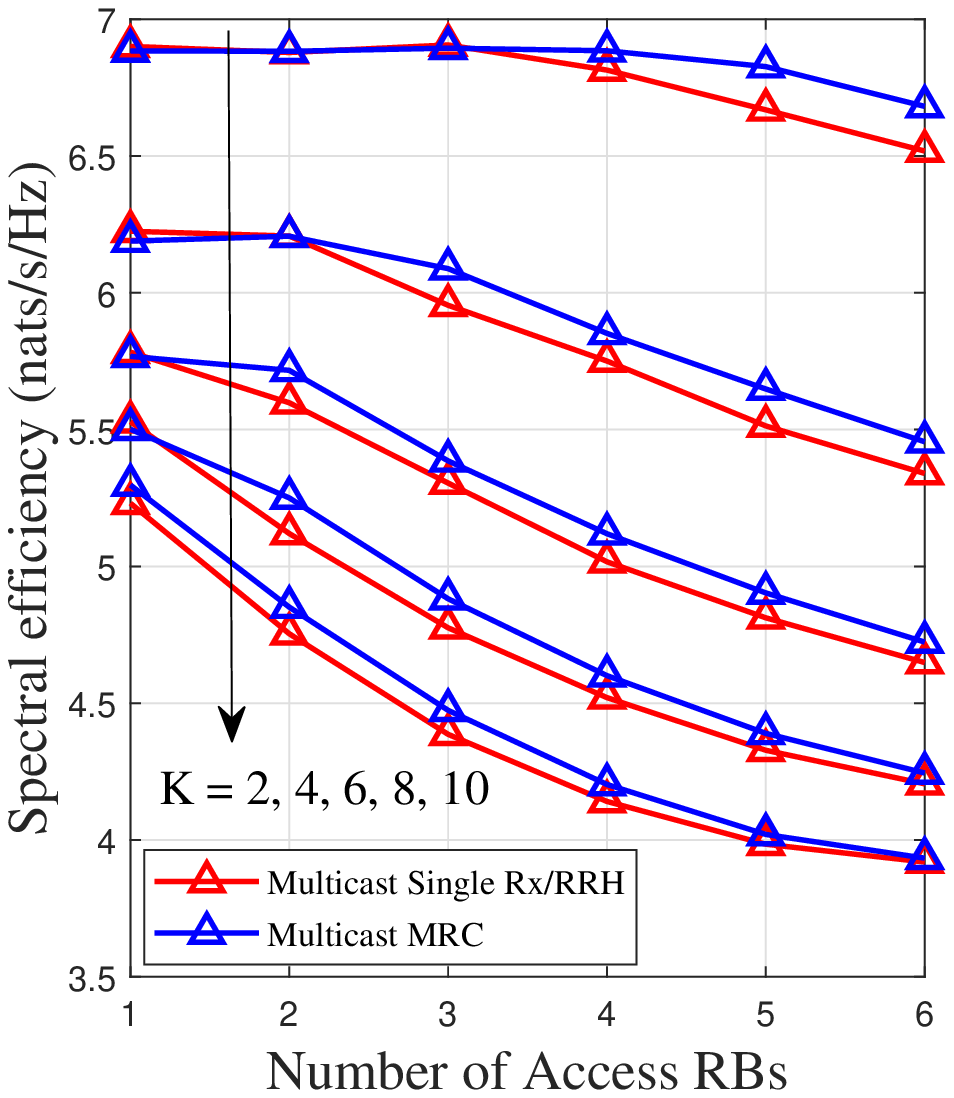}
		\vspace{-2em}
		\caption{Achieved spectral efficiency for different RB access configuration at $M=8$.}
		\label{fig:spectralEfficiency_multicast_sameNet_vsK_withM8}
	\end{minipage}%
	$\ $
	\begin{minipage}{.32\textwidth}
		\raggedleft
		\includegraphics[width=1\textwidth]{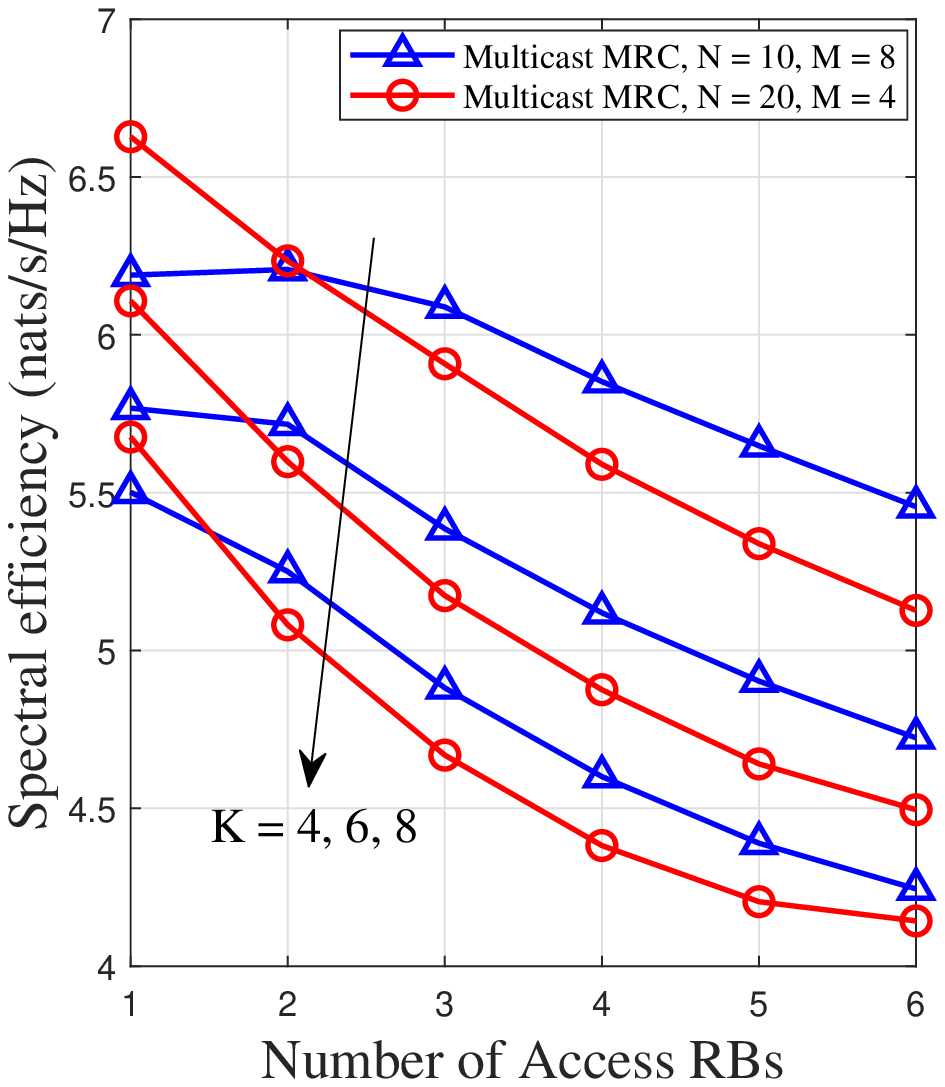}
		\vspace{-2em}
		\caption{Spectral Efficiency using multicast MRC fronthaul with different configurations.}
		\label{fig:MulticastMRC_N10M8_vs_N20M4}
	\end{minipage}%
	\vspace{-1em}
\end{figure}

In Figs.~\ref{fig:spectralEfficiency_multicast_sameNet} and~\ref{fig:spectralEfficiency_multicast_sameNet_vsK_withM8}, we plot the spectral efficiency per user \textit{on the access channel} as a function of $\omega$, the number of RBs allocated for the access channel, different number of RRH antennas, $M$, and number of users, $K$, hence showing different degrees of diversity and multiplexing tradeoff. We observe two main trends. First, as we reduce $\omega$, the rate saturates because the RRHs can be freely deployed in the network, so allocating more RBs on the access channel, i.e., increasing $\omega$ will not change the RRHs' locations and hence not change the \textit{spectral efficiency}. For example, in Fig.~\ref{fig:spectralEfficiency_multicast_sameNet_vsK_withM8} with $K=2$ we observe that using $\omega \le 5$ RBs allows for a free deployment of RRHs. Hence, under this network configuration $\omega$ should never be set less than $5$ because it wastes access channel resources, and it will allocate bandwidth to the fronthaul more than needed. The second trend, which can be seen at $\omega = 6$ RBs for $K=10$ in Fig.~\ref{fig:spectralEfficiency_multicast_sameNet_vsK_withM8}, is the right side of the curve where the fronthaul constraint becomes tight, i.e., the spectral efficiency deteriorates. 

In Fig.~\ref{fig:spectralEfficiency_multicast_sameNet}, to compare our results to a benchmark scheme, we also provide the performance resulting from placing the RRHs on a circle around the CU with as large a radius as possible while still respecting the statistical fronthaul constraint. We also constrain the radius of the circle to $2/3$ the distance between the CU and the cell edge on the $x$-axis. The results show that our optimization algorithm provides an $7\%$ of improvement in the spectral efficiency compared to this circular placement.

Our results suggest that a distributed network must be carefully designed; specifically, a key implication is that network operators must {choose} their bandwidth allocations carefully to retain the freedom to locate the RRHs. We note that sustaining multi-antenna communications at distributed nodes requires substantial resources be devoted to the fronthaul. Similarly, attempting to schedule a large number of users on the same time-frequency block leads to a very tight constraint on the fronthaul and, in turn, the access rate. To avoid deploying RRHs near their CUs, we need to reduce $K$, or, worse, allocate less spectrum to the access channel.

In Fig.~\ref{fig:MulticastMRC_N10M8_vs_N20M4} we plot the access channel spectral efficiency using multicast MRC fronthaul when we distribute the available antennas on more RRHs (a more distributed network). We compare the configurations $(N=10,\ M=8)$ and $(N=20,\ M=4)$ for different values of $K$ and $\omega$. In general, the results show that using a multicast fronthaul, distributing the antennas on more RRHs degrades performance. This is because multicasting becomes less effective as the number of receivers increases; furthermore, co-locating receive antennas improves receive diversity. The only exception is at very low allocated access channel bandwidth (e.g., at $\omega = 1$ in Fig.~\ref{fig:MulticastMRC_N10M8_vs_N20M4}) where, as explained earlier, the spectral efficiency on the access channel saturates. So, in such a case, there is less room to increase the spectral efficiency on the access channel, even if the multicast fronthaul is tighter. This observation is backed up with the saturation of the spectral efficiency at $\omega = 2$ RBs and $\omega = 1$ RB for $K = 4$ for the case of $(N = 20, M = 4)$. The saturation clearly shows that the access channel spectral efficiency cannot be further improved because the RRHs are already freely deployed (the fronthaul constraint is relatively loose).

\subsection{RRH Location Optimization with ZF Fronthaul}
\begin{figure}[t]
	\begin{minipage}{.46\textwidth}
		\centering
		\includegraphics[width=1\textwidth]{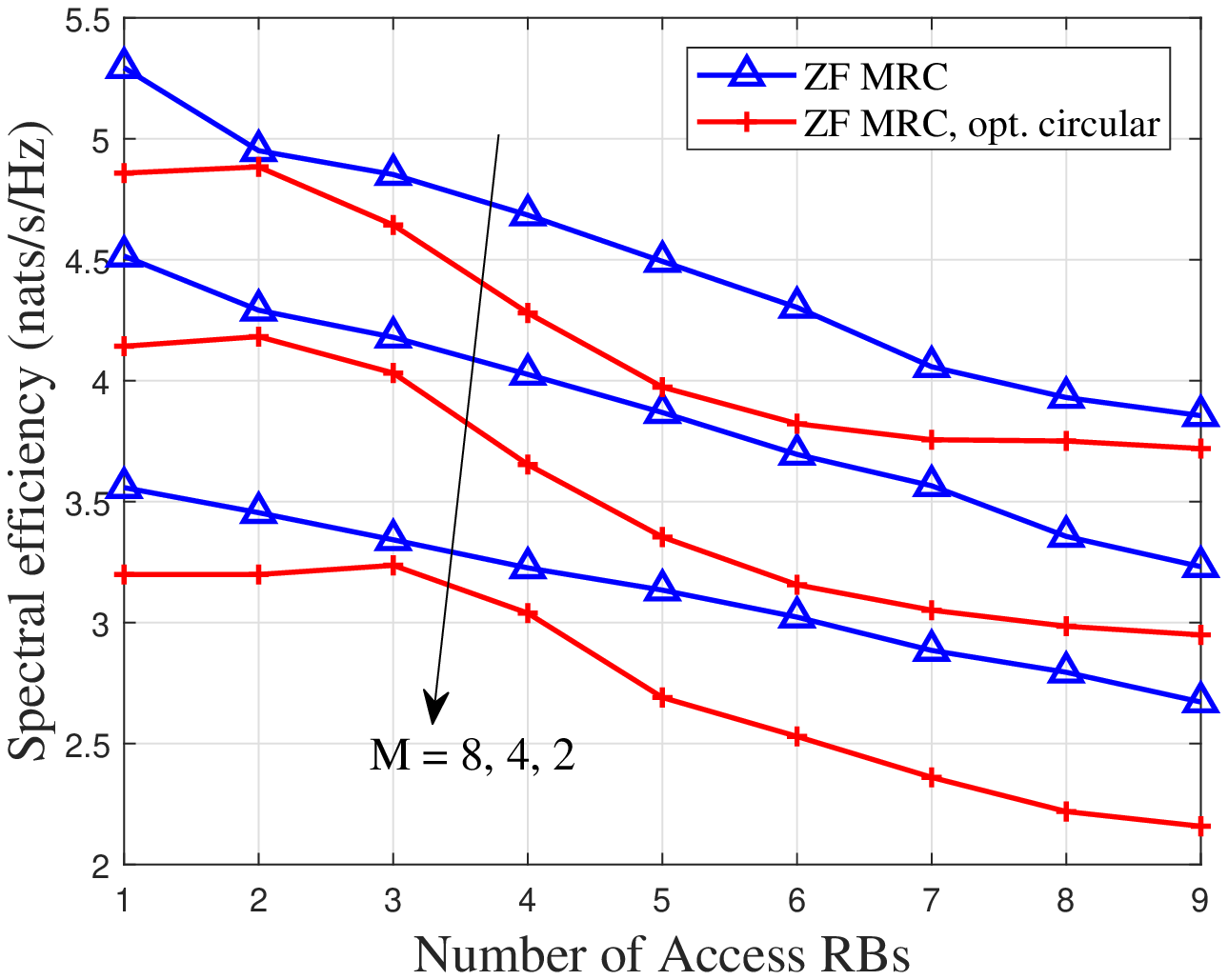}
		\vspace{-2em}
		\caption{Spectral Efficiency using ZF beamforming with MRC on the fronthaul.}
		\label{fig:SpectralEfficiency_ZFBF}
	\end{minipage}%
	\quad\quad
	\begin{minipage}{.46\textwidth}
		\centering
		\includegraphics[width=1\textwidth]{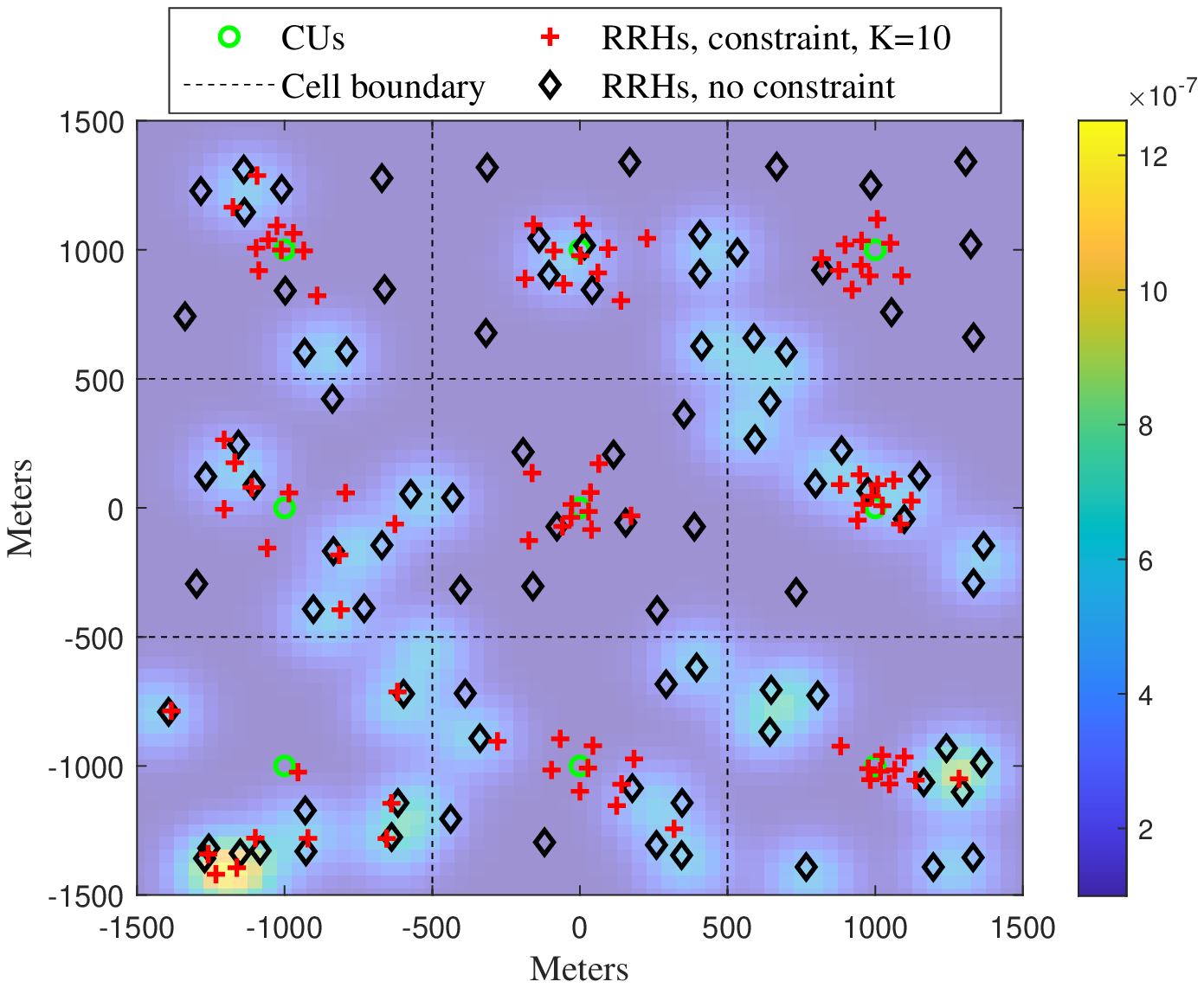}
		\vspace{-2em}
		\caption{RRHs deployed locations using ZF beamforming fronthaul, $M = 8,\ \omega = 4~\text{RBs}$.}
		\label{fig:deployZFBFMRC_M8RB4_M8RB1}
	\end{minipage}%
	\vspace{-1em}
\end{figure}

In Fig.~\ref{fig:SpectralEfficiency_ZFBF} we plot the spectral efficiency per user on the access channel when using ZF beamforming on the fronthaul. Fig.~\ref{fig:deployZFBFMRC_M8RB4_M8RB1} shows the corresponding deployment of RRHs without a fronthaul constraint and when $\omega = 4$ RBs. A comparison with the deployment in Fig.~\ref{fig:spectralEfficiency_multicast_sameNet} (multicast fronthaul) for the same configuration shows that using ZF beamforming in the fronthaul provides better performance and a relatively better deployment for the RRHs. Furthermore, Fig.~\ref{fig:SpectralEfficiency_ZFBF} quantifies the improved performance, for example, at $\omega = 4$ and $M = 8$, using a ZF beamforming fronthaul provides an efficiency of $4.7~{\rm nats/s/Hz}$ for each user, compared to $4.2~{\rm nats/s/Hz}$ for each user in the multicast case (as shown in Fig.~\ref{fig:spectralEfficiency_multicast_sameNet}), representing an improvement of 11\%.

\label{page:highlightedResults}If we compare the different bandwidth allocations and antenna configurations in both Figs.~\ref{fig:spectralEfficiency_multicast_sameNet} and~\ref{fig:SpectralEfficiency_ZFBF} while averaging the results, the wireless ZF fronthaul scheme with MRC provides an $8\%$ increase in the user's spectral efficiency compared to the multicast with MRC fronthaul. Moreover, our scheme provides a $11\%$ increase in the spectral efficiency per user compared to a circular placement that just barely meets the fronthaul constraint.

\begin{figure}[t]
	\begin{minipage}{.32\textwidth}
		\raggedright
		\includegraphics[width=1\textwidth]{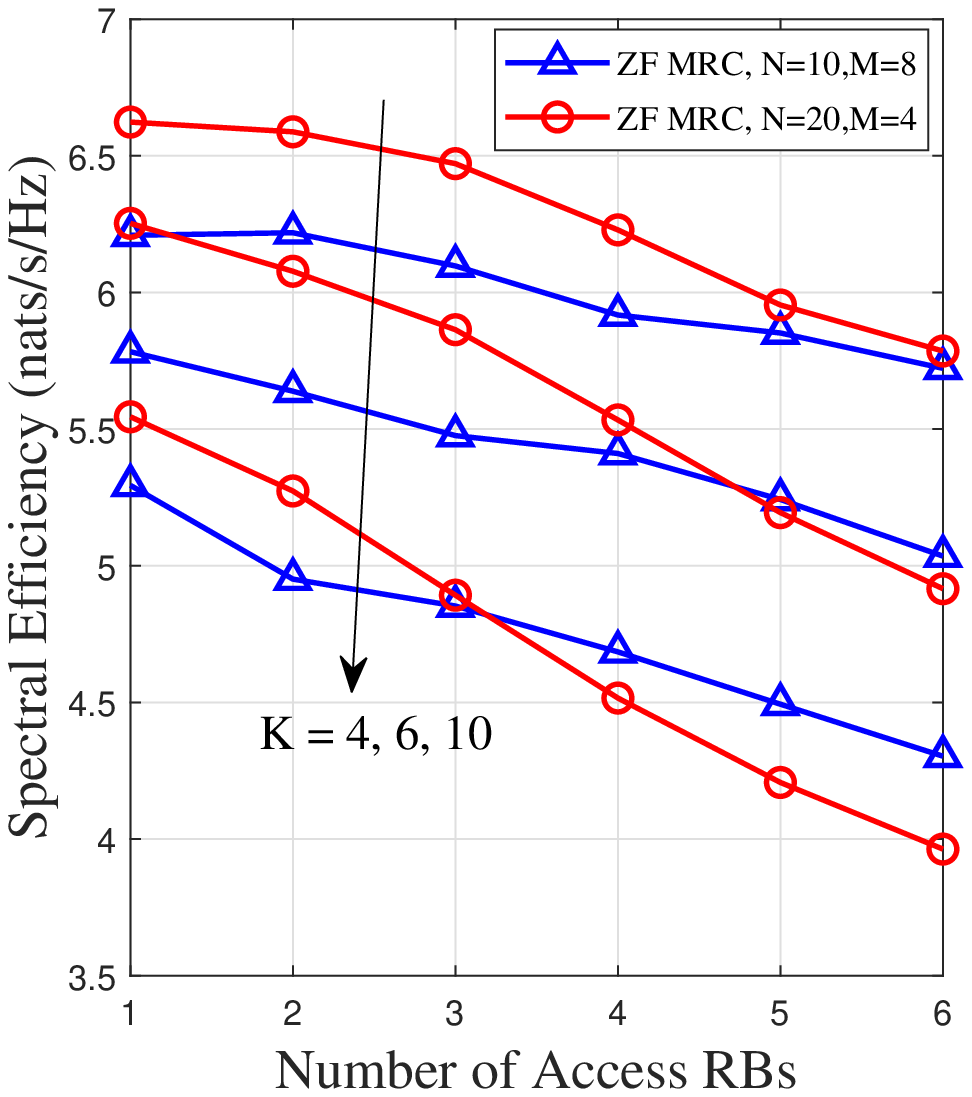}
		\vspace{-2em}
		\caption{Spectral Efficiency using ZF beamforming with MRC on the fronthaul.}
		\label{fig:ZFBF_N10M8_vs_N20M4_K4}
	\end{minipage}%
	$\ $
	\begin{minipage}{.32\textwidth}
		\centering
		\includegraphics[width=1\textwidth]{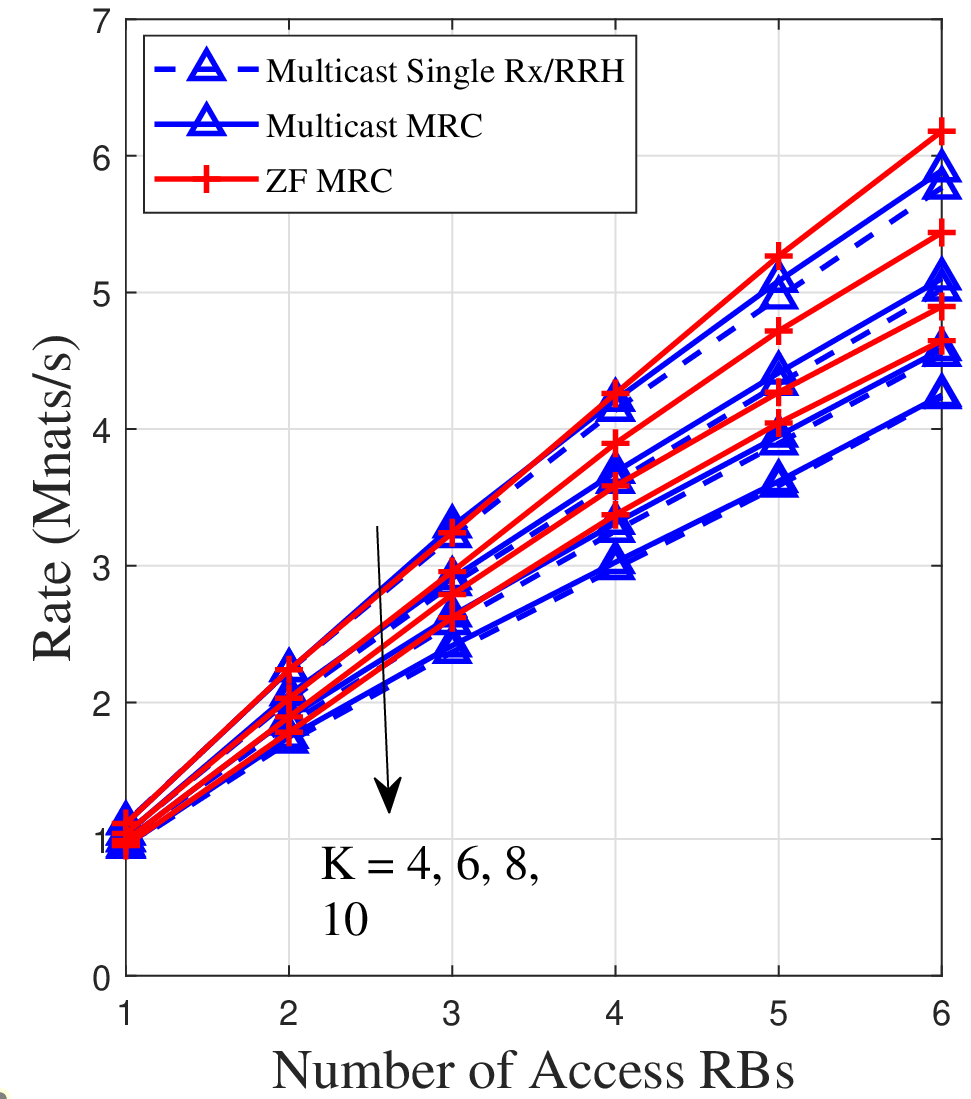}
		\vspace{-2em}
		\caption{Typical user rate at $N=10$, $M=8$.\\~}
		\label{fig:rate_ZFBF_multicast_samePlot_vsK_M8}
	\end{minipage}%
	$\ $
	\begin{minipage}{.33\textwidth}
		\raggedleft
		\includegraphics[width=1\textwidth]{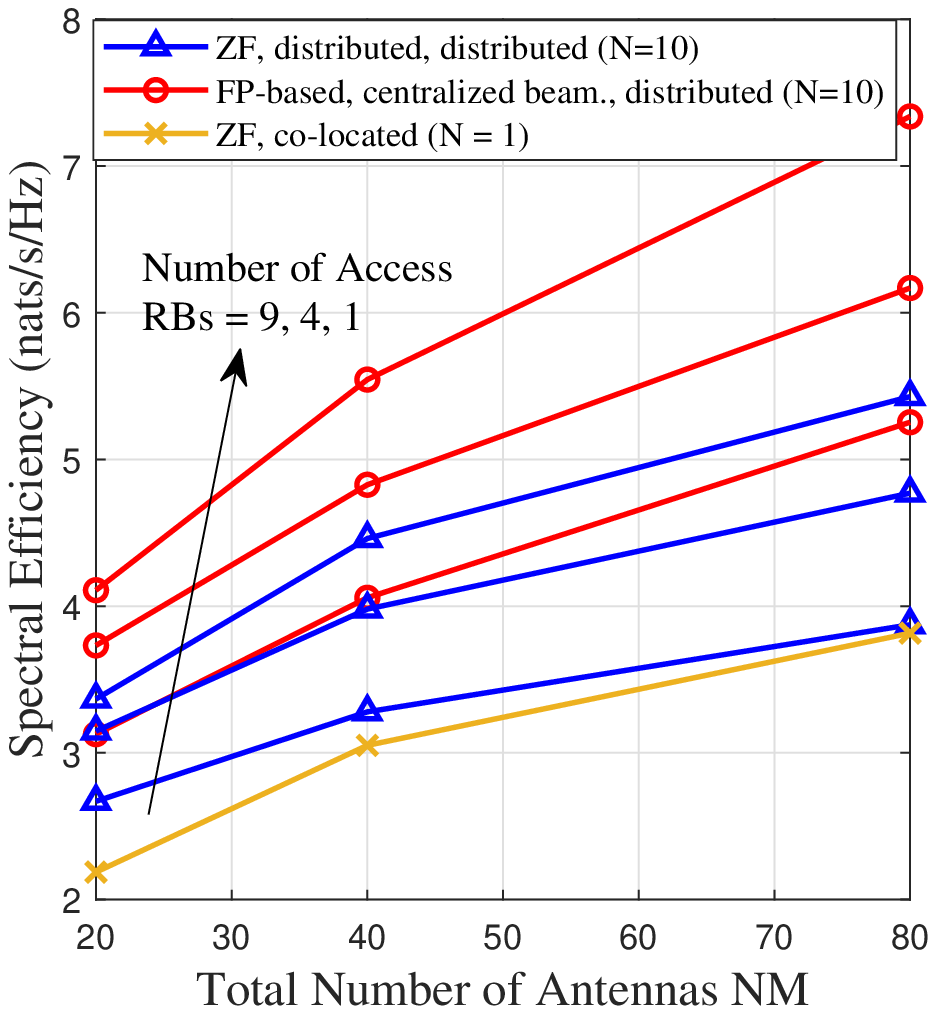}
		\vspace{-2em}
		\caption{Comparison between different scenarios for the access channel.}
		\label{fig:ZF_MMSE_N10_K10}
	\end{minipage}%
	\vspace{-1em}
\end{figure}

As a counterpart to Fig.~\ref{fig:MulticastMRC_N10M8_vs_N20M4}, in Fig.~\ref{fig:ZFBF_N10M8_vs_N20M4_K4}, we compare the performance of two configurations of 80 total antennas $(N=10,\ M=8)$ RRHs per cell vs a more distributed implementation corresponding to $(N=20,\ M=4)$. For most values of $\omega$, distributing the available antennas improves performance. This is because of the improved access channels (RRHs closer to users on average) outweighs the loss in diversity. Hence, for a ZF beamforming fronthaul, distributing the available antennas on more RRHs is the correct direction to boost network performance.

While Figs.~\ref{fig:spectralEfficiency_multicast_sameNet},~\ref{fig:spectralEfficiency_multicast_sameNet_vsK_withM8} and~\ref{fig:SpectralEfficiency_ZFBF} plot the spectral efficiency, in Fig.~\ref{fig:rate_ZFBF_multicast_samePlot_vsK_M8}, we plot the access rate per typical user, i.e., the spectral efficiency multiplied by the access channel bandwidth $\omega$ and the bandwidth per RB, as a function of the number of the RBs allocated for the access channel. As the plots show, the rates grow sublinearly with the ZF beamforming on the fronthaul providing the most freedom in deploying the RRHs. The results show that a rate gain is obtained from allocating more RBs on the access channel \emph{however, we emphasize that above $\omega = 6$ access channel RBs, the RRHs are essentially co-located with the CU, i.e., we do not have a distributed network}. Importantly, our results indicate that to deploy a wireless fronthaul solution with disjoint clustering and joint transmission requires considerable bandwidth to be available, so that we reap all the gains of coordinated distributed networks.

Finally, using Monte Carlo simulations, in Fig.~\ref{fig:ZF_MMSE_N10_K10} we plot the spectral efficiency when using ZF beamforming on the access channel and when using a numerically optimized beamforming technique~\cite{8314727} that is based on fractional programming (FP). The used FP approach is an advanced centralized optimization technique to construct the beamformers, i.e., the channels and auxiliary variables used to update the beamformers need to be executed at a single point in the network. We also plot the spectral efficiency when the antennas are co-located at a single point, i.e., the $NM$ antennas are co-located at a single RRH found at the cell center, thereby resembling a massive MIMO scenario. Although our paper does not optimize the beamformers, we plot these results to show the possible gap in the spectral efficiency when an advanced optimization technique is used. The results show that we may obtain an increase of $17\%-35\%$ in the spectral efficiency, which should be taken into consideration if such an advanced centralized beamforming technique is used. However, as noted earlier, using ZF provides closed-form expressions needed for the optimization of the locations of the RRHs, it also does not require centralized optimization for the beamformers and hence does not exhibit computational complexity and control plane load that may prevent implementation.

All in all, if a different beamforming scheme is used, the analysis needs to be performed using newly derived expressions for the mean and variance of the desired signal and the interference (as those derived in \eqref{eq:SigfronthaulMean_Multicast}-\eqref{eq:IntfronthaulVar_Multicast}, \eqref{eq:MultAntSigMean}-\eqref{eq:MultAntIntVar}, and \eqref{eq:ZFBFMeanSignal}-\eqref{eq:ZFBFIqnVariance}). If the RRHs are already deployed, then we can still control the fronthaul outage by allocating higher$\backslash$lower bandwidth to the fronthaul or through decreasing$\backslash$increasing the number of served users $K$. If the beamforming scheme does not allow a statistical analysis (e.g., beamformer is numerically optimized), a possible workaround would be to use the existing analysis but with an additional overhead for the argument of the CDF of the outage in the constraint (constraint~\eqref{eq:multicast_singleRxOpt} for multicast fronthaul and~\eqref{eq:Opt3_Constraint1} for zero forcing fronthaul). The overhead could relate to the difference in capacity provided when using the alternative numerical beamforming technique.

\section{Conclusion}\label{sec:conclusion}
In this paper, we have provided a statistical analysis of a wireless fronthaul for use in distributed networks. Specifically, we studied the effect of using a multicast or a ZF beamforming fronthaul when optimizing the deployment of RRHs. Importantly, we provide accurate closed-form expressions for the first two moments of the signal and interference, leading to a statistical characterization of the fronthaul. We then use this characterization to develop an optimization problem maximizing the user access rates while placing an outage constraint on the fronthaul.

Our results show that the fronthaul acts as a significant bottleneck on the deployment of RRHs. Specifically, unless considerable available bandwidth is devoted to the fronthaul, the RRHs must be placed very close to the CU. This, in turn, significantly reduces the achievable access rates. This bottleneck can be partially resolved by serving a low number of users on the same time-frequency resource block. Importantly, our characterization shows the trends in the network configuration and provides an optimized placement for the RRHs.


\appendices
\begingroup
\allowdisplaybreaks
\section{Derivation of \eqref{eq:SigfronthaulMean_Multicast}-\eqref{eq:IntfronthaulVar_Multicast}}
\label{appendix:meanVar_singleRXOutage} 
For the multicast fronthaul with a single receive antenna, $S^{(b)}_{qn}$ can be calculated as
\begin{align}
\mathbb{E}&\left\{S^{(b)}_{qn}\right\}
= 
p_{\rm c} \bar{\mu}_q^2 \bar{\ell}_q(x_{qn},y_{qn}) \mathbb{E}\bigg\{ \bigg|{\bf \bar{g}}_{q,qn}^H \sum_{n' = 1}^{N} \frac{{\bf \bar{g}}_{q,qn'}}{\sqrt{\bar{\ell}_q(x_{qn'},y_{qn'})}}\bigg|^2 \bigg\}
\nonumber \\
&
= p_{\rm c} \bar{\mu}_q^2 \bar{\ell}_q(x_{qn},y_{qn}) \mathbb{E}\big\{ \left|{\bf \bar{g}}_{q,qn}^H {\bf \bar{G}}_q {\bf a}_q\right|^2 \big\}
=
p_{\rm c} \bar{\mu}_q^2 \bar{\ell}_q(x_{qn},y_{qn}) \mathbb{E}\left\{ {\bf a}_q^H {\bf \bar{G}}_q^H {\bf \bar{g}}_{q,qn} {\bf \bar{g}}_{q,qn}^H {\bf \bar{G}}_q {\bf a}_q \right\}
\nonumber \\
&=
\resizebox{0.33\textwidth}{!}
{$\displaystyle
p_{\rm c} \bar{\mu}_q^2 \bar{\ell}_q(x_{qn},y_{qn}) 
\mathbb{E}\left\{
\text{{\bf tr}}\left\{ {\bf X}_{q,qn} {\bf A}_q \right\}
\right\}
$}
\stackrel{(a)}{\simeq}
\resizebox{0.51\textwidth}{!}
{$\displaystyle
p_{\rm c} \bar{\mu}_q^2 \bar{\ell}_q(x_{qn},y_{qn})\bigg( \frac{M_{\rm c}^2}{\bar{\ell}_q(x_{qn},y_{qn})} + \sum_{n' \ne n}^{N} \frac{M_{\rm c}}{\bar{\ell}_q(x_{qn'},y_{qn'})} \bigg)
$}
\end{align}
where ${\bf \bar{G}}_q= \left[{\bf \bar{g}}_{q,q1} \dots {\bf \bar{g}}_{q,qN} \right]\in \mathbb{C}^{M_{\rm c} \times N}$ and ${\bf a}_q = \Big[\frac{1}{\sqrt{\bar{\ell}_q(x_{q1},y_{q1})}} \dots \frac{1}{\sqrt{\bar{\ell}_q(x_{qN},y_{qN})}} \Big]^T \in \mathbb{R}^{N}$. Also, we define the following terms ${\bf X}_{q',qn} = {\bf \bar{G}}_{q'}^H {\bf \bar{g}}_{q',qn} {\bf \bar{g}}_{q',qn}^H {\bf \bar{G}}_{q'} \in \mathbb{C}^{N \times N}$ and ${\bf A}_q = {\bf a}_q {\bf a}_q ^H \in \mathbb{R}^{N \times N}$. The approximation in $(a)$ is very accurate due to the independence of the expressions definitions of $\mathbf{X}_{q,qn}$ and $\mathbf{A}_q$, and from the mean of $\bar{\mathbf{g}}_{q,qn} \bar{\mathbf{g}}_{q,qn}^H$.

For the interference, we have:
\begin{align}
\mathbb{E}\left\{I^{(b)}_{qn} \right\}
&= p_{\rm c} \mathbb{E}\left\{ \sum_{q' \ne q}^{Q} \bar{\mu}_{q'}^2 \bar{\ell}_{q'}(x_{qn},y_{qn}) \left|{\bf \bar{g}}_{q',qn}^H {\bf \bar{G}}_{q'} {\bf a}_{q'} \right|^2 \right\}
= p_{\rm c} \sum_{q' \ne q}^{Q} \bar{\mu}_{q'}^2 \bar{\ell}_{q'}(x_{qn},y_{qn}) 
\mathbb{E}\left\{
\text{\textbf{tr}}\left\{ {\bf X}_{q',qn} {\bf A}_{q'} \right\} 
\right\}
\nonumber \\
&= p_{\rm c} \sum_{q' \ne q}^{Q} \bar{\mu}_{q'}^2 \bar{\ell}_{q'}(x_{qn},y_{qn}) M_{\rm c} \sum_{n' = 1}^{N} \frac{1}{\bar{\ell}_{q'}(x_{q'n'},y_{q'n'})}
\end{align}
For the variance of $S^{(b)}_{qn}$, the proof starts by evaluating the variance for ${\bf \bar{g}}_{q,qn}^H {\bf \bar{g}}_{q,qn}$ at the diagonals which is $M_{\rm c}$. This leads us to the variance of the term ${\bf \bar{G}}_q^H {\bf \bar{g}}_{q,qn} {\bf \bar{g}}_{q,qn}^H \in \mathbb{C}^{N \times M_{\rm c}}$ which a matrix with $M_{\rm c}$ as each element except for the $n^\text{th}$ row which has $M_{\rm c}^2$ as its elements. In a similar fashion, this leads us to the variance of each entry in ${\bf X}_{q,qn} = {\bf \bar{G}}_q^H {\bf \bar{g}}_{q,qn} {\bf \bar{g}}_{q,qn}^H {\bf \bar{G}}_q \in \mathbb{C}^{N \times N}$, denoted as
\begin{align}
\mathtt{Var}\{{\bf X}_{q,qn}\} =
\footnotesize
\begin{bmatrix}
M_{\rm c}^2 & \dots & M_{\rm c}^2 & M_{\rm c}^3 & M_{\rm c}^2 & \dots & M_{\rm c}^2 \\
\vdots & \ddots & \vdots & \vdots & \vdots & \ddots & \vdots \\
M_{\rm c}^2 & \dots & M_{\rm c}^2 & M_{\rm c}^3 & M_{\rm c}^2 & \dots & M_{\rm c}^2 \\
M_{\rm c}^3 & \dots & M_{\rm c}^3 & 4M_{\rm c}^3 & M_{\rm c}^3 & \dots & M_{\rm c}^3 \\
M_{\rm c}^2 & \dots & M_{\rm c}^2 & M_{\rm c}^3 & M_{\rm c}^2 & \dots & M_{\rm c}^2 \\
\vdots & \ddots & \vdots & \vdots & \vdots & \ddots & \vdots \\
M_{\rm c}^2 & \dots & M_{\rm c}^2 & M_{\rm c}^3 & M_{\rm c}^2 & \dots & M_{\rm c}^2 \\
\end{bmatrix}
\end{align}
where the term $4M_{\rm c}^3$ is found at the $n^\mathrm{th}$ diagonal entry. We emphasize that this matrix denotes the variance of the individual entires in $\mathbf{X}_{q,qn}$. The variance of ${\bf a}_q^H {\bf \bar{G}}_q^H {\bf \bar{g}}_{q,qn} {\bf \bar{g}}_{q,qn}^H {\bf \bar{G}}_q {\bf a}_q \in \mathbb{C}$ is a linear combination of the path loss terms found in ${\bf a}_q$ and the matrix of variances of the entries ${\bf X}_q$ using $\mathtt{Var}\{{\bf a}_q^H {\bf X}_q {\bf a}_q\} = \mathtt{Var}\left\{\sum_{i} \sum_{j} \left[{\bf a}_q\right]_i \left[{\bf a}_q^H\right]_j \left[{\bf X}_q\right]_{ij}\right\}$ and $\mathtt{Var}\{aX\} = a^2 \mathtt{Var}\{X\}$. Hence:
\begin{align}
\mathtt{Var}\left\{S^{(b)}_{qn}\right\}
& =\left(p_{\rm c} \bar{\mu}_q^2 \bar{\ell}_q(x_{qn},y_{qn}) \right)^2 \mathtt{Var}\bigg\{ \bigg|{\bf \bar{g}}_{q,qn}^H \sum_{n' = 1}^{N} \frac{{\bf \bar{g}}_{q,qn'}}{\sqrt{\bar{\ell}_q(x_{qn'},y_{qn'})}}\bigg|^2 \bigg\}
\nonumber \\
&\ 
= \left(p_{\rm c} \bar{\mu}_q^2 \bar{\ell}_q(x_{qn},y_{qn}) \right)^2 \mathtt{Var}\left\{ {\bf a}_q^H {\bf \bar{G}}_q^H {\bf \bar{g}}_{q,qn} {\bf \bar{g}}_{q,qn}^H {\bf \bar{G}}_q {\bf a}_q \right\}
\nonumber\\
&\ 
=
\left(p_{\rm c} \bar{\mu}_q^2 \bar{\ell}_q(x_{qn},y_{qn}) \right)^2
\bigg( 4M_{\rm c}^3 \bigg(\frac{1}{\bar{\ell}_q(x_{qn},y_{qn})}\bigg)^2
\ 
+ \frac{M_{\rm c}^3}{\bar{\ell}_q(x_{qn},y_{qn})}\sum_{n' \ne n}^{N} \frac{1}{\bar{\ell}_q(x_{qn'},y_{qn'})}
\nonumber \\
&\quad 
+ 
\sum_{n' \ne n}^{N} \frac{1}{\bar{\ell}_q(x_{qn'},y_{qn'})} 
\bigg(\frac{M_{\rm c}^3}{\bar{\ell}_q(x_{qn},y_{qn})} + M_{\rm c}^2 \sum_{n'' \ne n}^{N} \frac{1}{\bar{\ell}_q(x_{qn''},y_{qn''})} \bigg)
\bigg)
\nonumber\\
& = 
\left(p_{\rm c} \bar{\mu}_q^2 \bar{\ell}_q(x_{qn},y_{qn}) \right)^2
\bigg( 4M_{\rm c}^3 \left(\frac{1}{\bar{\ell}_q(x_{qn},y_{qn})}\right)^2
+ 2\frac{M_{\rm c}^3}{\bar{\ell}_q(x_{qn},y_{qn})}\sum_{n' \ne n}^{N} \frac{1}{\bar{\ell}_q(x_{qn'},y_{qn'})}
\nonumber \\
& \quad
+ M_{\rm c}^2 \sum_{n' \ne n}^{N} \frac{1}{\bar{\ell}_q(x_{qn'},y_{qn'})} \sum_{n'' \ne n}^{N} \frac{1}{\bar{\ell}_q(x_{qn''},y_{qn''})}
\bigg)
\end{align}
Similarly, the variance for the interference $I^{(b)}_{qn}$ can be evaluated in a similar manner:
\begin{align}
&\mathtt{Var}\left\{I^{(b)}_{qn}\right\}
= p_{\rm c}^2 \mathtt{Var}\left\{ \sum_{q' \ne q}^{Q} \bar{\mu}_{q'}^2 \bar{\ell}_{q'}(x_{qn},y_{qn}) \left|{\bf \bar{g}}_{q',qn}^H {\bf \bar{G}}_{q'} {\bf a}_{q'} \right|^2 \right\}
\nonumber \\
&\ \stackrel{(a)}{=}
	\sum_{q' \ne q}^{Q} \left(p_{\rm c} \bar{\mu}_{q'}^2 \bar{\ell}_{q'}(x_{qn},y_{qn}) \right)^2 \mathtt{Var}\left\{ {\bf a}_{q'}^H {\bf \bar{G}}_{q'}^H {\bf \bar{g}}_{q',qn} {\bf \bar{g}}_{q',qn}^H {\bf \bar{G}}_{q'} {\bf a}_{q'}  \right\}
\nonumber \\
&\ =
\sum_{q' \ne q}^{Q} \left(p_{\rm c} \bar{\mu}_{q'}^2 \bar{\ell}_{q'}(x_{qn},y_{qn}) \right)^2
M_{\rm c}^2 \sum_{n' \ne n}^{N} \frac{1}{\bar{\ell}_{q'}(x_{q'n'},y_{q'n'})} \sum_{n'' \ne n}^{N} \frac{1}{\bar{\ell}_{q'}(x_{qn''},y_{qn''})}
\nonumber \\
&\ =
\sum_{q' \ne q}^{Q} \bigg(p_{\rm c} \bar{\mu}_{q'}^2 \bar{\ell}_{q'}(x_{qn},y_{qn}) M_{\rm c} \sum_{n' = 1}^{N} \frac{1}{\bar{\ell}_{q'}(x_{q'n'},y_{q'n'})} \bigg)^2
\end{align}
where $(a)$ follows since, for $X$ and $Y$ independent, $\mathtt{Var}\left\{X+Y\right\} = \mathtt{Var}\left\{X\right\} + \mathtt{Var}\left\{Y\right\}$.

\section{Derivation of \eqref{eq:MultAntSigMean}-\eqref{eq:MultAntIntVar}}\label{appendix:multicast_multiRX_outage}
For the case with multiple receive antennas and using MRC, the mean of the desired signal $S^{(b)}_{qn}$ can be written as
\vspace{-0.5em}
\begin{align}
\mathbb{E}&\left\{S^{(b)}_{qn}\right\}=
p_{\rm c} \bar{\mu}_q^2 \bar{\ell}_q(x_{qn},y_{qn}) \mathbb{E}\bigg\{ \bigg|{\bf b}_{qn}^H {\bf \bar{G}}_{q,qn}^H \sum_{n' = 1}^{N} \frac{{\bf \bar{G}}_{q,qn'}{\bf b}_{qn}}{\sqrt{\bar{\ell}_q(x_{qn'},y_{qn'})}}\bigg|^2 \bigg\}
\nonumber \\
&
= p_{\rm c} \bar{\mu}_q^2 \bar{\ell}_q(x_{qn},y_{qn}) \mathbb{E}\bigg\{ \big|{\bf b}_{qn}^H {\bf \bar{G}}_{q,qn}^H {\bm \Upsilon}_q {\bf a}_q\big|^2 \bigg\}
=
p_{\rm c} \bar{\mu}_q^2 \bar{\ell}_q(x_{qn},y_{qn}) \mathbb{E}\left\{ {\bf a}_q^H {\bm \Upsilon}_q^H {\bf \bar{G}}_{q,qn} {\bf b}_{qn} {\bf b}_{qn}^H {\bf \bar{G}}_{q,qn}^H {\bm \Upsilon}_q {\bf a}_q \right\}
\nonumber \\
&\stackrel{(a)}{=} p_{\rm c} \bar{\mu}_q^2 \bar{\ell}_q(x_{qn},y_{qn}) \left(\mathbb{E}\left\{\sigma_{q,qn}^{(G)}\right\}\right)^2 \left(\sigma_1^{(G)}\right)^2
\mathbb{E}\left\{ {\bf a}_q^H {\bf U}_{q1}^H \left[{\bf U}_{q,qn}\right]_{.1} \left[{\bf U}_{q,qn}\right]_{.1}^H {\bf U}_{q1} {\bf a}_q \right\}
\nonumber \\
&= p_{\rm c} \bar{\mu}_q^2 \bar{\ell}_q(x_{qn},y_{qn}) \left(\mathbb{E}\left\{\sigma_{q,qn}^{(G)}\right\}\right)^4
\mathbb{E}\left\{
\text{\textbf{tr}}\left\{ {\bf X}_{q,qn} {\bf A}_q \right\}
\right\}
\nonumber \\
& = p_{\rm c} \bar{\mu}_q^2 \bar{\ell}_q(x_{qn},y_{qn})
\bigg( \frac{1}{\bar{\ell}_q(x_{qn},y_{qn})}
+ \frac{1}{M_{\rm c}^2\sqrt{\bar{\ell}_q(x_{qn},y_{qn})}} \sum_{n' \ne n}^{N} \frac{1}{\bar{\ell}_q(x_{qn'},y_{qn'})}
\nonumber \\
&\ \ 
\hspace*{0.2in} +
\sum_{n' \ne n}^{N} \bigg( \frac{1}{M_{\rm c} \bar{\ell}_q(x_{qn'},y_{qn'})} + \frac{1}{M_{\rm c}^2 \sqrt{\bar{\ell}_q(x_{qn},y_{qn}) \bar{\ell}_q(x_{qn'},y_{qn'})}} \bigg)
\bigg)
\end{align}
where ${\bm \Upsilon}_q = [\left({\bf \bar{G}}_{q,q1} {\bf b}_{q1}\right)\ \left({\bf \bar{G}}_{q,q2} {\bf b}_{q2}\right)\ \dots\ \left({\bf \bar{G}}_{q,qN} {\bf b}_{qN}\right)] \in \mathbb{C}^{M_{\rm c} \times N}$ and ${\bf \bar{G}}_{q,qn} \in \mathbb{C}^{M_{\rm c} \times M}$ is the small-scale fading between CU $q$ and RRH $n \in \mathcal{D}_q$. The step in $(a)$ follows from the fact that ${\bf \bar{G}}_{q,qn} {\bf b}_{qn} = \sigma_1^{(G)} \left[{\bf U}_{q,qn}\right]_{.1}$, where $\left[{\bf U}_{q,qn}\right]_{.1}$ is the first column of the unitary matrix ${\bf U}_{q,qn}$ obtained from the SVD of the small-scale fading ${\bf \bar{G}}_{q,qn}$. Moreover, ${\bf U}_{q1} = \left[\left[{\bf U}_{q,q1}\right]_{.1}\ \left[{\bf U}_{q,q2}\right]_{.1}\ \dots\ \left[{\bf U}_{q,qN}\right]_{.1}\right]$. At last, ${\bf X}_{q,qn}={\bf U}_{q1}^H \left[{\bf U}_{q,qn}\right]_{.1} \left[{\bf U}_{q,qn}\right]_{.1}^H {\bf U}_{q1}$ and ${\bf A}_q = {\bf a}_q {\bf a}_q^H$. The mean and the variance of the interference can be derived in a similar fashion as in Appendix~\ref{appendix:meanVar_singleRXOutage} and we do not present them to preserve space.


\endgroup

\ifCLASSOPTIONcaptionsoff
\newpage
\fi

\footnotesize
\bibliography{Bk_References}
\bibliographystyle{ieeetr}

\end{document}